\documentclass[a4paper,12pt]{article}
\pdfoutput=1 
\usepackage{jheppub} 
\usepackage[T1]{fontenc}
\usepackage{graphicx}
\usepackage{bm}
\usepackage{comment}
\usepackage{subcaption}

\title{Swampland Constraints on No-Boundary Quantum Cosmology}

\author[a]{Hiroki Matsui}
\author[b]{Takahiro Terada}

\affiliation[a]{Theory Center, IPNS, KEK, 1-1 Oho, 
Tsukuba, Ibaraki 305-0801, Japan}
\affiliation[b]{Center for Theoretical Physics of the Universe, 
\\ Institute for Basic Science (IBS), Daejeon, 34126, Korea}

\emailAdd{matshiro@post.kek.jp}
\emailAdd{takahiro@ibs.re.kr}

\abstract{The Hartle-Hawking no-boundary proposal describes the quantum creation of the universe.  To have a non-negligible probability to obtain a classical expanding universe, eternal inflation is required, which is severely constrained by Swampland conjectures such as the refined de Sitter conjecture and the distance conjecture.  We discuss this issue in detail and demonstrate the incompatibility.  We show that the dimensionless parameters in the refined de Sitter conjecture should be bounded from above by a positive power of the scalar potential to realize the classical expanding universe.  In other words, the probability of the classical expanding universe is extremely small under the Swampland conjectures unless the parameters are much smaller than unity.   If they are order unity, on the other hand, the saddle-point solution itself ceases to exist implying a genuinely quantum universe.  }

\begin{document} 
\thispagestyle{empty}
\begin{flushright}
{\small CTPU-PTC-20-15\\
 KEK-TH-2221\\
 KEK-Cosmo-256}\\
\end{flushright}
\maketitle
\flushbottom

\section{Introduction}

The standard description of inflation is based on the semiclassical assumption 
that spacetime can be described by classical background 
with small quantum fluctuations, 
which is highly consistent with cosmological observations. 
However, it has difficulty with
the initial conditions with the spacetime singularity
at the very moment of the birth of the universe~\cite{Borde:2001nh} and cannot explain how the universe was born~\cite{Vilenkin:1982de}. 
To discuss the initial state of the universe and 
go beyond the standard paradigm of inflation,
the spacetime should be quantized, and the universe should be 
described by the paradigm based on quantum gravity i.e., quantum cosmology.

One of the focuses in quantum cosmology is the wave function (functional) of the universe, which describes the state of the universe. 
The state is specified by some prescriptions, such as choices of boundary conditions 
and the contour of the integral, for the Wheeler-DeWitt equation~\cite{DeWitt:1967yk} or the path integral in quantum gravity (which are formally equivalent).  There are several proposals for assigning such prescriptions: 
Hartle-Hawking no-boundary proposal~\cite{Hawking:1981gb, 
Hartle:1983ai, Hawking:1983hj} and the Vilenkin tunneling
proposal~\cite{Vilenkin:1984wp, Vilenkin:1986cy, Vilenkin:1987kf}.  
The former (latter) is usually formulated by the Euclidean
(Lorentzian) path integral.  In addition, there is Linde's proposal~\cite{Linde:1983cm, Linde:1984ir} closely related to them.  See Refs.~\cite{Linde:1984ir, Linde:2005ht, Vilenkin:1998dn, Vilenkin:2002ev} for the relations among them and other related work. 
These describe the creation of the homogeneous and isotropic universe from ``nothing'', namely from the universe with the geometry of zero sizes.

One of the largest differences between the cosmological predictions of the Hartle-Hawking no-boundary proposal and of the Vilenkin's and Linde's proposals is that the probability of creation of the universe with the scalar potential $V(\phi) (>0)$ is proportional to $\exp ( \pm 24 \pi^2 / V(\phi))$ where $+$ and $-$ correspond to the Hartle-Hawking no-boundary proposal and  Vilenkin's and Linde's proposals, respectively,\footnote{\label{fn:controversy}
Recently, Feldbrugge et al.~\cite{Feldbrugge:2017kzv} proposed a systematic reformulation of the no-boundary proposal which  predicts  the probability $\propto \exp ( - 24 \pi^2 /V)$ with the {\em opposite} sign of the exponent compared with the original no-boundary proposal, effectively unifying it with the Vilenkin's proposal.  Moreover, the resultant probability distribution of field perturbations is such that larger fluctuations are probabilistically favored
~\cite{Feldbrugge:2017fcc}.  If this is true, 
the no-boundary proposal as well as the tunneling one is to be discarded.   However, based also on the Picard-Lefschetz theory, Diaz Dorronsoro et al.~\cite{DiazDorronsoro:2017hti} reached a completely different conclusion, i.e., they found the traditional sign and suppressed fluctuations.  Their subsequent publications (Refs.~\cite{Feldbrugge:2017mbc, Feldbrugge:2018gin} and Refs.~\cite{DiazDorronsoro:2018wro, Halliwell:2018ejl, Janssen:2019sex}) continue to disagree, so the situation is controversial.  Note that the wave function predicted by Feldbrugge et al.~is Green's function rather than a solution to the Wheeler-DeWitt equation.
See also related discussions~\cite{Vilenkin:2018dch, Vilenkin:2018oja, Bojowald:2018gdt, DiTucci:2018fdg, DiTucci:2019dji, DiTucci:2019bui}.
} in the limit of the vanishing kinetic energy of the scalar field $\phi$. The difference corresponds to the difference of the imposed boundary conditions or prescriptions of the path integral.  The sign in the expression of the probability implies that Linde's and Vilenkin's wave functions predict the creation of the universe with the largest possible $V$, while the no-boundary wave function predicts that with the smallest possible $V$.

In the standard path integral approach, a state is approximated by the sum of relevant saddle-point solutions which satisfy appropriate boundary conditions.  
The state is said to be classical when the phase of the wave function of the universe changes more rapidly than the absolute value.  A continuously expanding solution easily becomes classical, and 
such a solution is associated with the subsequent classical history of the universe.  In this way, the wave function of the universe represents a probability distribution of classical histories of the universe. 
Indeed, the classicality condition tends to be satisfied 
when the universe experiences the inflationary (or the ekpyrotic) phase~\cite{Battarra:2014kga, Lehners:2015sia}. 
In the Vilenkin's and Linde's proposals, a larger value of $V$ is predicted, so it is not difficult to have a long
period of inflation. On the other hand, the no-boundary proposal predicts a smaller value of $V$, so it is nontrivial to have a sufficiently long period of inflation and satisfy the classicality. Hartle et al.~\cite{Hartle:2008ng} proposed to consider the probability under the condition that the
entire universe contains a patch similar to our observed universe, and such a conditional probability entails a volume-weight factor, which significantly enhances the probability to have a long period of inflation.  We revisit the issue of short inflation in the 
no-boundary proposal and  examine whether it can be solved by 
the volume-weight factor.

So far, we described an aspect of quantum cosmology which provides the quantum state of the universe, but we also need to specify the theory (Lagrangian).  We work with the Einstein gravity (plus a scalar field).   
In general, it should  be regarded as an effective field theory (EFT) valid up to the Planck scale, which will be ultraviolet-completed by a consistent theory of quantum gravity such as string theory. 
According to the Swampland conjectures~\cite{Vafa:2005ui, Ooguri:2006in}, any EFT 
consistent with quantum gravity  is believed to satisfy some nontrivial conditions, and EFTs incompatible with such conditions are said to belong to the Swampland rather than string Landscape (see reviews~\cite{Brennan:2017rbf, Palti:2019pca}).

For instance, a refined version~\cite{Klaewer:2016kiy, Baume:2016psm} of the Swampland distance conjecture~\cite{Ooguri:2006in} tells us that the masses of an infinite tower of particles scale as $\exp (- d \Delta \phi)$ 
 when any scalar field moves a distance $\Delta \phi (\gtrsim M_\text{P})$ where $d$ is an $\mathcal{O}(1)$ constant and $M_\text{P} (\equiv 1)$ is the reduced Planck mass. Below the cutoff scale $\Lambda$ of the EFT, there appears a huge number of particle species $N_\text{s} \sim \Lambda /(m e^{-d \Delta \phi})$ where $m$ is the initial mass of the lightest of the tower particles, which would be at most the Planck scale.  Such a large number of states lowers the cutoff  exponentially, $\Lambda < M_\text{P} / \sqrt{N_\text{s}}$~\cite{Dvali:2007hz}, so 
the EFT breaks down when $\Lambda \lesssim H$ where $H$ is the Hubble parameter, which is a rather conservative estimate (see Ref.~\cite{Palti:2019pca} for more details).  This
restricts the maximum possible field variation in the EFT~\cite{Scalisi:2018eaz}, 
 \begin{align}
 \Delta\phi \lesssim (3/d) M _{\rm P }  \log (M_\text{P} / H) \; . \label{SDC}
\end{align}
 The right-hand side becomes $(32/d) M_\text{P}$ when we take the Planck upper bound on the Hubble parameter~\cite{Akrami:2018odb}.

Another example is the refined version~\cite{Ooguri:2018wrx, Garg:2018reu} of the de Sitter (dS) Swampland Conjecture~\cite{Obied:2018sgi} (see also Refs.~\cite{Dvali:2018fqu, Dvali:2018jhn}). 
It  states that the low-energy effective potential $V$,
consistent with quantum gravity, must satisfy the following conditions,
\begin{align}\label{swampland}
 |V^\prime| \geq & c V \; ,  &  
&\mbox{or} & V^{\prime\prime} \leq -c^\prime V \; ,
\end{align}
where $|V^\prime|$ is the norm of the gradient of $V$
on the scalar manifold, $V''$ is the smallest eigenvalue of the second derivative of $V$, and $c$ and $c'$ are positive constants presumably not extremely smaller than $\mathcal{O}(1)$. The original version of the conjecture does not accept the second inequality, while the refined version accepts it. Alternative refinements are discussed in Refs.~\cite{Andriot:2018wzk, Andriot:2018mav}. 
These bounds~\eqref{swampland} forbid the de Sitter solutions, and slow-roll inflation is also in tension with the conjecture depending on the size of  $c$ and $c'$.  If they are $\mathcal{O}(1)$, single-field slow-roll inflation is forbidden or at least  incompatible with observations~\cite{Agrawal:2018own, Achucarro:2018vey, Garg:2018reu, Kinney:2018nny, Brahma:2018hrd, Das:2018hqy, Fukuda:2018haz,Ashoorioon:2018sqb}.   
A simple possibility for the resolution of this tension between slow-roll inflation and the de Sitter conjecture is to assume $c$ and $c'$ are somewhat smaller than $\mathcal{O}(1)$, say, $\mathcal{O}(10^{-1})$, or $\mathcal{O}(10^{-2})$.\footnote{ \label{fn:beyond_SFSR} 
Alternative possibilities include multi-field dynamics~\cite{Palti:2019pca, Achucarro:2018vey}, excited initial conditions~\cite{Brahma:2018hrd, Ashoorioon:2018sqb}, and warm inflation~\cite{Das:2019hto, Brandenberger:2020oav}.
} 
We have to comment that the status of the refined dS conjecture is much less established (see, e.g., Refs.~\cite{Palti:2019pca, Akrami:2018ylq}) than other Swampland conjectures.  We study its implications on the quantum cosmology assuming its validity.  Note, however, that problems of stable dS spacetime have been discussed in broader contexts~\cite{Tsamis:1996qq, ArkaniHamed:2007ky, Polyakov:2007mm, Dvali:2013eja, Dvali:2017eba}.

In the present paper, we discuss whether the quantum cosmology is consistent with these Swampland conjectures. 
In the Linde's and Vilenkin's proposals, there are no difficulties in classicalization and inflation if $c$ and $c'$ are somewhat smaller than $\mathcal{O}(1)$ since larger $V$ makes inflation be 
realized more easily, so we do not consider these cases further. Instead, we refer the reader to Ref.~\cite{Brahma:2020cpy} for discussions on the tunneling proposal in light of the trans-Planckian censorship conjecture (TCC)~\cite{Bedroya:2019snp, Bedroya:2019tba}. 
We note that the TCC was recently questioned as a Swampland criterion in Ref.~\cite{Saito:2019tkc} and significantly modified in Refs.~\cite{Seo:2019wsh, Cai:2019dzj}, so we do not consider it in this paper.\footnote{
Scalar weak gravity conjecture (scalar WGC)~\cite{Palti:2017elp} and its variants~\cite{Gonzalo:2019gjp, Heidenreich:2019zkl} are another class of Swampland conjectures which may be relevant.  However, the presence of many variants implies there is no clear guiding principle.  Also, some versions of the scalar WGC are known to be in phenomenological tension with the refined dS conjecture~\cite{Shirai:2019tgr}.  Therefore, we do not consider them in this paper.  
}

In the no-boundary proposal, on the other hand, the situation is more nontrivial.
The naive estimate predicts the creation of the universe with an extremely small $V$, so we need to consider the conditional probability with the volume-weight factor proposed by 
Hartle et al.~\cite{Hartle:2008ng} in the first place. 
However, we will show that this simple solution  is incompatible with the distance conjecture and/or the refined de Sitter conjecture.  This is because a sufficient amount of the volume factor requires significantly 
long-term inflation, i.e., eternal inflation, which is severely constrained by the Swampland conjectures~\cite{Matsui:2018bsy, Dimopoulos:2018upl, Kinney:2018kew, 
Brahma:2019iyy}.

This paper is organized as follows. 
In section~\ref{sec:no-boundary}, we review the Hartle-Hawking no-boundary proposal, in which the wave function of the universe is introduced as a Euclidean path integral.  The issue of the small probability for a long period of inflation and its presumed solution based on the conditional probability are explained in more detail. 
In section~\ref{sec:Swampland}, we discuss whether the no-boundary proposal is consistent with the refined dS conjecture and the distance conjecture.  The rough analytic argument is presented, which is followed by numerical analyses.  
Section~\ref{sec:Conclusion} is devoted to discussion and conclusions. 
Appendix~\ref{sec:generic_slow-roll} contains our analyses on the saddle-point solution for generic slow-roll potentials.

\section{No-boundary Quantum Cosmology}
\label{sec:no-boundary}

In this section, we review the no-boundary proposal. 
The no-boundary proposal is a prescription to give an initial state (or equivalently the initial wave function) of the universe.\footnote{An alternative interpretation is presented in Ref.~\cite{Linde:2005ht}.}  A prerequisite is that the universe has a closed geometry (but see Refs.~\cite{Coule:1999wg, Linde:2004nz, Linde:2017pwt}). 
The main idea is that  the wave function with some field values should be given by a Euclidean path integral~\cite{Hawking:1981gb, Hartle:1983ai, Hawking:1983hj} over any smooth compact four-dimensional spacetime geometry that have a three-dimensional spacelike boundary which reproduces the given field values.  That is, there is no need to impose the boundary condition at the moment of creation of the universe explicitly, but analogous conditions are introduced just as some requirement on the smoothness of the integrated geometries.  This is why it is called the ``no-boundary'' proposal.

The integral in general is complicated, but it significantly simplifies under the assumption of the minisuperspace approximation, where homogeneity and isotropy are assumed.
In the minisuperspace approximation, the scale factor $b = b(t)$ and the scalar field $\chi = \chi(t)$ are homogeneous where we reserve the notation $a$ and $\phi$ for the path-integration variables. 
The no-boundary wave function of the universe is defined (up to the normalization constant) by
\begin{equation}\label{w-function}
\Psi( b, \chi) = \int_{\cal{C}} \delta a \, \delta \phi \, 
\exp\left(-S_\text{E} [g_{\mu\nu},\phi]/\hbar \right)\;,
\end{equation}
where 
 $g_{\mu\nu} = \ell^2 (d \tau^2 + a^2 (\tau) d\Omega^2_3 )$ is the Friedmann-Lema\^{i}tre-Robertson-Walker (FLRW) metric with $a$, $\tau$, $\ell$, and $d\Omega_3^2$ being the scale factor, complex time (explained below), a length parameter, and the element of the three-sphere, respectively, $\phi$ is a scalar matter field, 
and $S_\text{E}$ is the Euclidean action,
\begin{align}
S_\text{E} =& \frac{1}{2}\int \text{d}^4 x  \sqrt{g}R - i \int \text{d}^3 x  \sqrt{h} K + \int \text{d}^4 x \sqrt{g} \left(  \frac{1}{2}g^{\mu\nu}\partial_\mu \phi \partial_\nu \phi + V(\phi) \right) \nonumber \\ 
=& 2\pi^2 \ell^3 \int \text{d}\tau \,  a^3 \left( -3 \left( \frac{\text{d}a}{\text{d}\tau} \right)^2 - \frac{3}{(\ell a)^2} + \frac{1}{2} \left(  \frac{\text{d}\phi}{\text{d}\tau}  \right)^2 + V(\phi) \right) ,
\end{align}
where the second term in the first line is the Gibbons-Hawking-York boundary term~\cite{York:1972sj, Gibbons:1976ue} with $h$ and $K$ denoting the determinant of the induced metric and the trace of the extrinsic curvature, respectively, on the boundary hypersurface. 
 $\ell$ is taken as the reduced Planck length in this section, but a different normalization is used in our numerical computations. 
The integration is carried out along the configuration paths $\cal{C}$ 
which starts with the no-boundary condition 
as we will see later and ends at the Lorentzian point where $d\tau^2 = - dt^2$. 
Since we smoothly interpolate the geometry between the Euclidean spacetime and the Lorentzian spacetime at the boundary time $t$, we have to consider the analytic continuation of time as well as the field variables $a$ and $\phi$. Thus, $a(\tau)$ and $\phi(\tau)$ have complex values in general, but its boundary values at time $t$, i.e., $b$ and $\chi$ must be real.

The path integral in eq.~\eqref{w-function} 
can be approximated by the saddle point approximation, and the 
no-boundary wave function takes the form,
\begin{align}
\Psi( b, \chi) \approx \exp\{- \left( S_\text{E} ^\text{R}(b,\chi) +iS_\text{E} ^\text{I}(b,\chi) \right) /\hbar\}  \;. \label{Psi_saddle-point}
\end{align}
Here and hereafter, we use a shorthand notation representing the real and the imaginary part 
by a superscript R and I, respectively.  
The right-hand side should be a sum over relevant\footnote{
Which saddle-point solutions are relevant is actually a nontrivial question tightly related to the controversy mentioned in Footnote~\ref{fn:controversy}.  In this paper, we take the traditional Hartle-Hawking saddle-point.}  saddle-point solutions when there are such multiple solutions.
In this way, the Lorentzian histories are approximately
described by the complex Euclidean instanton, 
$S_\text{E}(b, \chi)$, which is also called a fuzzy instanton since the distinction between the Lorentzian and the Euclidean is not always clear.

\subsection{Equations of motion and no-boundary conditions}

With the complex FLRW metric,
the Euclidean equations of motion are
\begin{align} 
&\left( \frac{\text{d} a }{\text{d} \tau} \right)^2  -1  - \frac{a ^2}{3} \Biggl[ \frac{1}{2} \left( \frac{\text{d} \phi }{\text{d} \tau} \right)^2 - V( \phi)  \Biggr] = 0 \;, \\
&\frac{\text{d}^2 a}{\text{d} \tau^2} + \frac{a}{3} \Biggl[\left( \frac{\text{d} \phi }{\text{d} \tau} \right)^2+ V( \phi)  \Biggr] = 0 \;,\\
&\frac{\text{d}^2 \phi }{\text{d} \tau^2}  + 3 \frac{ 1}{a}  \frac{\text{d} a }{\text{d} \tau} \frac{\text{d} \phi }{\text{d} \tau}- V' ( \phi)  = 0 \;,  
\end{align}
where  $\; ^{\prime} \equiv \text{d}/\text{d} \phi$. 
 The on-shell Euclidean action reads,
\begin{equation} \label{eq:complexAction}
S_\text{E}( b, \chi)_\textrm{on-shell} = 4\pi ^2 \int_{\cal{T}} d \tau \left( -3a + a ^3V( \phi) \right) \;,
\end{equation}
where $\mathcal{T}$ is the complex integration path for the time $\tau$. Note that it does not explicitly depend on the kinetic energy of $\phi$.

We evaluate the Euclidean path integral in the semiclassical approximation 
by solving the equations of motion along the complex contour $\cal{T}$.
The integral depends only on the endpoint of the complex contour $\mathcal{T}$, at which $a$ and $\phi$ reduce to $b$ and $\chi$, respectively,   
as long as there are no singularities or branch cuts in the complex plane.
Thus, we can take any contour for $\tau = x + i y$ with $x$ and $y$ real.

The no-boundary proposal imposes a regularity condition
at the ``South Pole'' ($\tau = 0$), 
\begin{gather}
a(\tau = 0) = 0 \; ,\qquad
\frac{\text{d}a}{\text{d}\tau}(\tau = 0) = 1 \; ,\qquad
\frac{\text{d}\phi}{\text{d}\tau}(\tau = 0) = 0 \; ,
\end{gather}
and the following condition at the endpoint ($x =X$, $y = Y$), 
\begin{gather}
a(\tau = X + i Y) = b \; ,\qquad
\phi(\tau = X + i Y) = \chi \; ,
\end{gather}
where ($b$, $\chi$) are real values at the end point.
Note that we must choose a complex scalar field value at the South Pole 
to satisfy the Wentzel-Kramers-Brillouin (WKB) classicality conditions~\cite{Hartle:2008ng},
\begin{align}
\phi(\tau = 0) \equiv \phi_\text{SP} (\in \mathbb{C}) \; .  
\end{align}
It is known that the fine-tuning of one parameter (such as the phase for a given absolute value and the imaginary part for a given real part) for the complex scalar field $\phi_\text{SP}$
at the South Pole is necessary to obtain the classical Lorentzian universe with  
high probabilities~\cite{Hartle:2008ng}.

\subsection{Classicality conditions}

To realize classical Lorentzian histories 
with high probabilities, we impose the WKB classicality condition on the wave function.
Inserting the saddle-point wave function~\eqref{Psi_saddle-point} into the Wheeler-DeWitt equation 
and expanding it in powers of $\hbar$, we obtain~\cite{Lyons:1992ua, Hartle:2008ng}
\begin{align}
\frac{1}{2}\left(\nabla S_\text{E}^\text{R}\right)^2 - i  \nabla S_\text{E}^\text{R}\cdot \nabla S_\text{E}^\text{I} - \frac{1}{2}\left(\nabla S_\text{E}^\text{I}\right)^2+U=  0 \; , 
\end{align}
at the leading order, where $U \equiv a^3 V / 3$ is the potential of the Wheeler-DeWitt equation, 
and the Laplacian is defined concerning the metric of the minisuperspace model:  $\nabla^2 \equiv G^{AB}\partial_A \partial_B $ ($A, B = b, \chi$) with $G^{bb}= - \frac{1}{12\pi^2 b}$, $G^{\chi\chi} = \frac{1}{2 \pi^2 b^3}$, and $G^{b\chi}=G^{\chi b}=0$.\footnote{We take a different normalization convention than that of Ref.~\cite{Lehners:2015sia}} 
When the amplitude of the wave function  changes more slowly than the phase does, i.e., when the WKB classicality condition 
\begin{align}
\frac{| \partial _{b} S_\text{E} ^\text{R}|}{| \partial _{b} S_\text{E} ^\text{I}|}\ll & 1 \; , & \text{and}& & 
\frac{| \partial _{\chi} S_\text{E} ^\text{R}|}{| \partial _{\chi} S_\text{E} ^\text{I}|}\ll & 1 \; , \label{WKB}
\end{align}
is satisfied, the Lorentzian Hamilton-Jacobi equation is approximately realized,
\begin{align}
-\frac{1}{2}\left(\nabla S_\text{E}^\text{I}\right)^2+U=0 \; ,
\end{align}
which is analogous to the WKB method in quantum mechanics.
In this case, the universe behaves classically.
As discussed in Appendix~\ref{sec:generic_slow-roll}, it turns out that the classicality condition is met when $b$ is large and $\chi$ is such that the potential is large.

Once classicality is achieved, $b$ and $\chi$ behave classically.  For such values of $b$ and $\chi$, the relative probability density, or the probability density up to the overall normalization factor,  of the Lorentzian histories in the leading semiclassical approximation is given by~\cite{Hartle:2008ng},
\begin{align}\label{probability}
P( \chi)  =& | \Psi ( b, \chi ) |^2  \nabla _{b} S_\text{E}^\text{I} (b, \chi)  \nonumber \\
\approx &  \exp[-2 S_\text{E} ^\text{R}(b,\chi)] 
\nonumber  \\
= & \exp \left( -8\pi ^2 {\rm Re}\left[\int_{\cal{T}} d \tau \left( -3a + 
a ^3V(\phi)  \right)\right]\right) \;,
\end{align}
where a constant $b$ is assumed in this expression. 
In the second (approximate) equality, the factor $\nabla _{b} S_\text{E}^\text{I}$ and higher-order contributions in $\hbar$ are neglected. 
In the first equality, the right-hand side is independent of $b$ provided that the classicality condition is satisfied~\cite{Hartle:2008ng}.  The dependence on $b$ in the last expression, if any, appears due to the approximations.   
 The condition of classicality and the existence of an inflationary epoch are tightly correlated with each other~\cite{Battarra:2014kga, Lehners:2015sia}.  
The universe must eventually be dominated by the potential energy to 
satisfy the WKB classicality condition and realize the classical Lorentzian histories. 
In other words, the kinetic term and the spatial curvature should be subdominant, and the slow-roll condition should be satisfied.

\subsection{Example}\label{sec:example}

\begin{figure}[thb!] 
\begin{center}
\includegraphics[width=0.42\columnwidth]{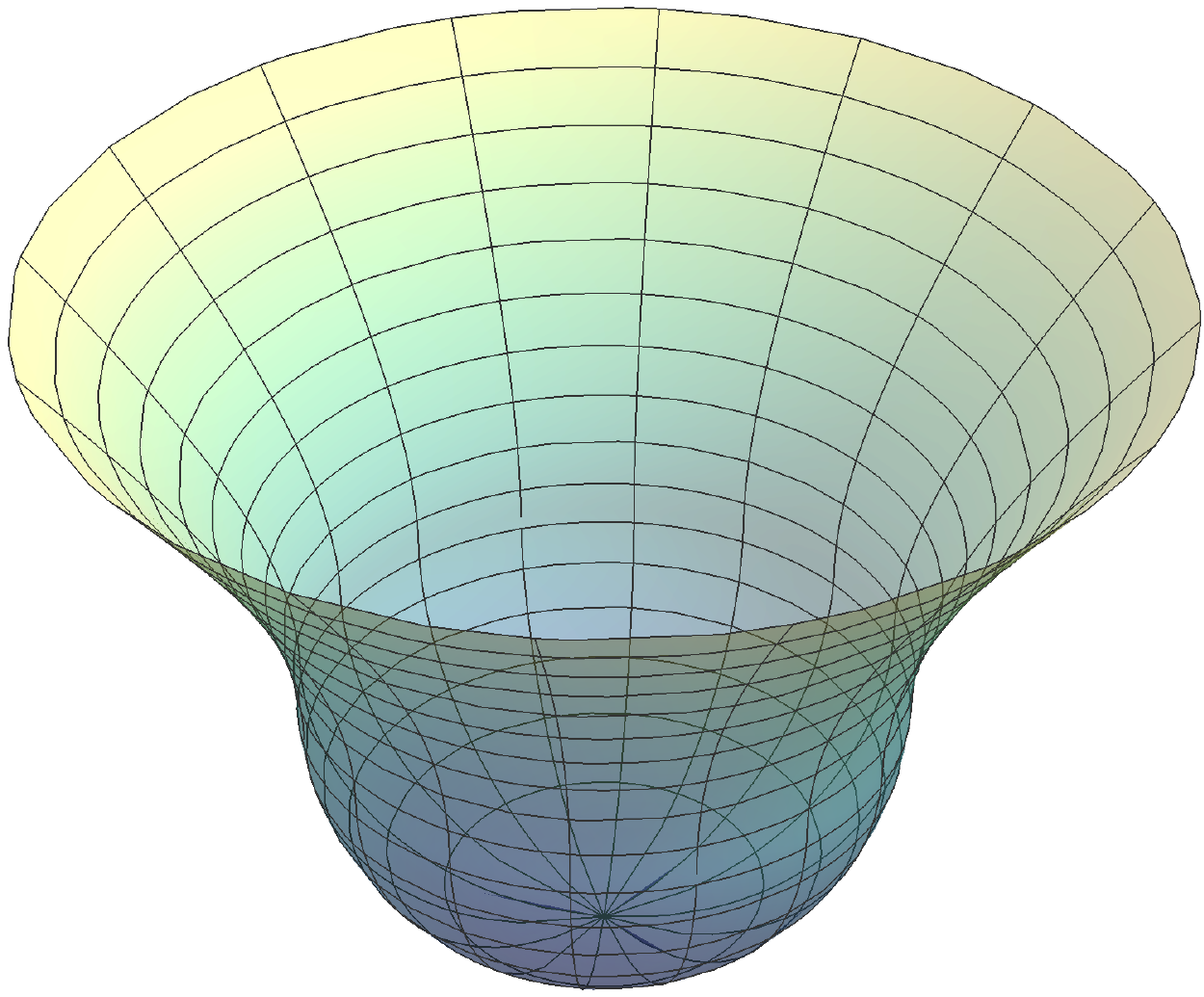}
\caption{Schematic view of the creation of de Sitter universe from nothing embedded in a higher dimensional flat Euclidean spacetime.  The half-sphere in the Euclidean part is smoothly connected to the half hyperboloid in the Lorentzian part.}
\label{fig:universe_nucleation}
\end{center}
\end{figure}

To develop intuitions, it is instructive to consider the simplest example in which the scale factor is the only degree of freedom.  The cosmological constant $\Lambda$ is introduced instead of the scalar field potential.  The Euclidean Friedmann equation is solved as 
\begin{align}
a(\tau) = & \sqrt{\frac{3}{ \Lambda}} \sin \left( \sqrt{\frac{\Lambda}{3}} \tau \right) \;  .
\end{align}
Note that the imaginary part of $a(\tau)$ vanishes on the lines specified by $\tau^\text{R}$ such that $\cos \sqrt{\frac{\Lambda}{3}}\tau^\text{R} = 0$.  The first positive solution corresponds to the maximum of the sine function.  From this point, the (real) scale factor increases exponentially in the imaginary (i.e. Lorentzian) direction.  To see this, it is convenient to parametrize $\tau = (\pi/2) \sqrt{3/\Lambda} + i t$, so that 
\begin{align}
a(t) = \sqrt{\frac{3}{\Lambda}} \cosh \left( \sqrt{\frac{\Lambda}{3}} t \right).
\end{align}
Along the contour, the scale factor is always real.  This exponential growth corresponds to the de Sitter expansion.
These solutions are smoothly connected at $\tau =  \sqrt{3/\Lambda} \, \pi/2$ ($t=0$) and describe the creation of the Lorentzian de Sitter universe from nothing through a transient Euclidean half-sphere as depicted in Fig.~\ref{fig:universe_nucleation}.\footnote{
The vertical axis is the coordinate $x^0$ in the embedding spacetime which is related to $\tau$ and $t$ as $\sin (\sqrt{\Lambda/3} \tau - \pi/2) = x^0$ and $\sinh (\sqrt{\Lambda / 3} t) = x^0$. The radius in the horizontal plane represents the size of the universe $\sqrt{\Lambda/3} a(x^0)= \sqrt{ 1 + \text{sign}(x^0) (x^0)^2}$ where sign$(x^0)$ denotes the sign of $x^0$.
} 

In the Euclidean regime (on the real axis with $\tau^\text{R} < \sqrt{3/\Lambda} \, \pi /2 $), the Euclidean action is real,
\begin{align}
S_\text{E} 
 = &  \frac{12 \pi^2 }{ \Lambda}  \Biggl[  \cos^3 \left( \sqrt{\frac{ \Lambda }{3}} \tau\right) -1 \Biggr] \; ,
\end{align}
so the classicality condition is not satisfied.
In the Lorentzian regime, it is written as 
\begin{align}
S_\text{E} = & \frac{12\pi^2}{\Lambda} \Biggl[ - 1 + i \sinh \left( \sqrt{\frac{\Lambda}{3}} t \right)^3 \Biggr] \; .
\end{align}
The real part is constant, while the imaginary part changes rapidly.  Hence, the classicality condition is satisfied in this regime.

With the addition of the scalar field and its potential, the creation of the chaotic inflationary universe in the context of the no-boundary proposal has been well investigated, e.g., in Refs.~\cite{Lyons:1992ua, Hartle:2007gi, Hartle:2008ng}.  In these cases, there is usually no contour along which both $a$ and $\phi$ are always real, so the boundary of the quantum and the classical regimes, as well as the Euclidean and Lorentzian regimes becomes nontrivial. 
In Appendix~\ref{sec:generic_slow-roll}, we generalize the analyses of Refs.~\cite{Lyons:1992ua,Hartle:2007gi, Hartle:2008ng} to the case of generic slow-roll potentials and evaluate the Euclidean action and the classicality condition.

\subsection{Issue of Small Probability for Classical Inflationary Universe}  \label{subsec:small-probability-issue}

The relative nucleation probability density (simply ``probability'' for short) of the universe for generic classes of slow-roll scalar potentials is approximately given by
\begin{align}\label{pprobability}
P(\chi)  \approx  \exp[-2 S_\text{E} ^\text{R}(b,\chi)]
\approx { \exp \left( \text{Re} \, \frac{24 \pi ^2}{ V(\phi_\text{SP}) } \right)  }  \;.
\end{align}
This is derived in Appendix~\ref{sec:generic_slow-roll}, but we can reproduce it heuristically by replacing $\Lambda$ in the previous subsection with $V(\phi_\text{SP})$ and by extracting its real part.  
The probability is exponentially larger for smaller values of the potential. 
This implies that it is unlikely to have sufficiently long inflation, and it is also nontrivial to have a classical universe.
Even if the universe becomes classical, it will collapse shortly after it is nucleated and classicalized.\footnote{
It was pointed out in Ref.~\cite{Matsui:2019ygj} that it is in principle possible to increase $V$  in the contraction phase of the universe while suppressing the rapid growth of other energy components by fine-tuning of the functional form of the potential $V$ and initial conditions.  If the universe subsequently experiences a non-singular bounce due to positive spatial curvature as in Refs.~\cite{Matsui:2019ygj, Sloan:2019jyl}, which naturally leads to an inflationary phase and does not violate the null energy condition, the resultant cosmology is viable.  Note that the required positive sign of the spatial curvature is consistent with the no-boundary proposal. 
}
Thus, the no-boundary proposal is in tension with the inflationary universe. 

This may be a serious issue.
Substituting the present dark energy, the above exponential factor becomes $\exp (3 \times10^{122})$~\cite{Page:2006hr}.
It implies that the probability of nucleating the macroscopic empty universe in the present vacuum is extremely larger than the probability of nucleating the inflationary universe followed by the standard hot big-bang.
We note, however, that this is not necessarily a problem when one adopts the anthropic arguments or the environmental selection effects.  Nevertheless, 
it would be much better if the probability to obtain realistic observation-compatible universes is high.

For a more quantitative discussion, one may introduce the total probability of the inflationary universe and  that of the non-inflationary universe where classicality is realized but inflation does not happen. (In the latter case, the universe tend to collapse.)  They are given by~\cite{Page:1997vc, Hwang:2013nja},
\begin{align}\label{inf-probability}
P_{\rm inf} \equiv &
\int _{ { \chi  }_{ \rm inf} }^{ { \chi  }_\textrm{cut-off} }{ P(\chi ) } d\chi  \; , & 
P_\textrm{non-inf} \equiv & 
\int _{ { \chi  }_{ \rm cl} }^{ { \chi  }_\textrm{inf} }{ P(\chi ) } d\chi ,
\end{align}
where ${ \chi  }_{ \rm cl}$ is the critical field value satisfying  the classicality conditions,  ${ \chi  }_{ \rm inf}$ is the critical value  for slow-roll inflation, for which we take $\epsilon (\chi_\text{inf}) = 1$ for definiteness, and ${ \chi  }_\textrm{cut-off}$ is the cut-off field value for the probability integral. 
 In these definitions, it is implicitly assumed that $\chi$ rolls down in the negative direction as a convention (so that $\chi_\text{cl} < \chi_\text{inf} < \chi_\text{cut-off}$), but the generalization is obvious.  
Practically, the numerical integration as defined above is not necessarily convenient since there is typically a large hierarchy of the integrand in the integration domain.  Our following discussion is based on $P(\chi)$ with the understanding that $P_\text{inf}$ and $P_\text{non-inf}$ are typically completely dominated by an endpoint of the integration.

\subsubsection{Conditional Probability and Volume Weight}
\label{subsec:Hartle-Hawking-Hertog}
In general, the no-boundary proposal provides us with the extremely suppressed probabilities for nucleation of the inflationary universes, and Eq.~\eqref{probability} expresses the probabilities of the entire cosmological histories. 
However, most of the universes in the probability distribution will collapse or perhaps goes back to the quantum regime, so they will not evolve into universes like ours.  
Hartle, Hawking, and Hertog~\cite{Hartle:2007gi, Hartle:2008ng}  suggested to consider the probability distribution under the condition that the whole universe contains at least one region where our observed universe exists. (See also a closely related discussion based on the anthropic argument~\cite{Page:1997vc}.)  Although it would be a complicated task to precisely evaluate such a conditional probability, it should be proportional to its spatial volume. 
Thus, they introduced the following volume-weighted relative probability density,
\begin{align}
P_{\textrm{volume-weight}} \equiv & e^{3 \mathcal{N}_e} P(\chi)  \nonumber \\
= & e^{3 \mathcal{N}_e} \exp[-2 S_\text{E} ^\text{R}(b,\chi)]\;,  \label{volume-weighted_probability}
\end{align}
where we take the e-folding number $\mathcal{N}_e$ to be that during inflation for definiteness, 
\begin{equation}\label{e-folding}
\mathcal{N}_e = \int_{ \chi_\text{inf}}^{\chi} \frac{V(\phi)}{V'(\phi)} d\phi \; ,
\end{equation}
where we assumed slow-roll. 

To understand when the volume factor becomes effective enough to compensate the otherwise small probability, we consider the minimum of the volume-weighted probability. 
The extremality condition is
\begin{align}\label{minimum-probability}
&\frac { \partial P_{\textrm{volume-weight}} }{ \partial { \chi  } } = 0\, \nonumber \\
&\Longrightarrow \frac { \partial }{ \partial { \phi  } } 
\left( \frac{24\pi^2}{V(\phi)} + 3 \int\frac{V(\phi)}{V'(\phi)} d\phi \right) = 0\nonumber \\ & \Longrightarrow \quad\frac{V'(\phi)^2}{V(\phi)^3} = \frac{1}{8\pi^2}\;.
\end{align}
This approximately coincides with the critical value of the slope of the potential for eternal inflation~\cite{Page:1997vc} since the condition for eternal inflation is~\cite{Barenboim:2016mmw, Rudelius:2019cfh}
\begin{align}
\frac{V'(\phi)^2}{V(\phi)^3} \lesssim & \frac{1}{12 \pi^2} & \text{and}& &  \frac{V''}{V}\gtrsim& - 3 \;, \label{eternity_condition}
\end{align}
where the first condition ensures the dominance of quantum fluctuations $\delta_\text{q} \phi$ over the classical motion $\delta_\text{c} \phi$, and the second condition, which is relatively less severe, ensures the dominance of the volume expansion rate over the decay rate from the field region supporting the eternal inflation.\footnote{
The numerical coefficient in the first inequality in Eq.~\eqref{eternity_condition} is derived here from a simple criterion $\delta_\text{q} \phi  \gtrsim \delta_\text{c} \phi$ where $\delta_\text{q} \phi = H/(2\pi)$ and $\delta_\text{c} \phi = |\dot{\phi}| / H$, so there is an $\mathcal{O}(1)$ difference with the analysis  based on the Fokker-Planck equation in Ref.~\cite{Rudelius:2019cfh}.  The second criterion is derived for a quadratic hilltop potential~\cite{Rudelius:2019cfh}, but we expect that a similar criterion applies to the second derivative of more generic scalar potentials unless higher derivatives are hierarchically large.  See also Ref.~\cite{Barenboim:2016mmw} for a related discussion on this point.  Note also that the second inequality (and its underlying requiment) does not play a major role for the plateau-type potential, which we mainly study.
} 
The volume-weighted probability begins to grow around the point where eternal inflation takes place.
Hence, the volume-weighted proposal can realize the inflationary universe, and it predicts eternal inflation.   This conditional probability is the relevant probability for a given observer who infers the past of the universe he or she lives in.

When we consider an energy scale much smaller than the Planck scale ($24\pi^2/V \gg 1$), the required e-folding number becomes extraordinarily large, and it will enter the eternal inflation regime.  Then, the e-folding number, i.e., how long inflation continues in a given part of the universe, is dominantly determined by the stochastic quantum fluctuations rather than the classical slow-roll dynamics.   In the eternal inflation regime, it is not transparent for us to clearly define an average (i.e., to define an appropriate measure) of the e-folding.   In the next section, we will take two approaches.  In the first approach, we follow Refs.~\cite{Hartle:2007gi, Hartle:2008ng} and extrapolate the above slow-roll formula~\eqref{e-folding}.  In the second approach, instead of concretely evaluating the e-folding number, we simply assume that a sufficiently large e-folding number required to produce observation-compatible universes with non-negligible probability is realized in some part of the eternally inflating universe.

Before moving on to the main part of the paper, we briefly comment on another potential way to enhance the probability of the classical inflationary universe by considering a multifield flat potential.\footnote{
Other approaches have beeen discussed in Refs.~\cite{Hwang:2011mp, Hwang:2012zj, Hwang:2014vba}.}
The probability of an inflationary universe $P_\text{inf}$ in eq.~\eqref{inf-probability} increases when the integration domain is large as discussed in Ref.~\cite{Hwang:2013nja}, and it is efficient when the integrand $P(\chi) \sim \exp (24 \pi^2 /V)$ is nearly constant in the integration region.  That is, if the potential is asymptotically flat, the probability increases.  However, Swampland conjectures such as the distance conjecture severely constrain the available field space length. One may argue that the problem is ameliorated when we consider a multifield flat potential.  Suppose that the multiple scalar fields form an irreducible representation of a(n approximate) global symmetry group such as O($N$) with $N \gg 1$.  In this case, the probability increases as $\text{Vol}(\mathbb{B}^N) \chi_\text{cutoff}^N$ where $\text{Vol}(\mathbb{B}^N) \sim ( 2\pi e / N)^{N/2}$ is the volume of the $N$-dimensional unit ball. The probability increases exponentially in terms of $N$ only if the field distance is larger than $N /(2 \pi e) (\gg  1 )$.  Thus, even if we consider the multifield generalization, this possibility is, at best, inefficient compared to the volume factor originating from the conditional probability discussed above.

\section{No-boundary Proposal vs Swampland Conjectures}
\label{sec:Swampland}
In this section, we discuss whether the no-boundary proposal
is consistent with the (refined) de Sitter Swampland conjecture~\eqref{swampland} and the distance conjecture~\eqref{SDC}. 
We will show that there is significant tension between them. 

\subsection{Setup and assumptions}

An immediate consequence of the distance and dS conjectures is that we cannot take an infinitely extended plateau potential since it violates both conjectures.  The plateau-type feature is nevertheless favored by the observations~\cite{Akrami:2018odb}, so we consider an approximate plateau potential.  That is, we consider  potentials with a  finite plateau and a broken shift symmetry such as 
\begin{align} 
V = V_0 \, \Biggl[ \left(\tanh \frac{\phi}{\sqrt{6 \alpha }} \right)^2 + \varepsilon  \cosh  \frac{\phi}{\sqrt{6 \alpha }} \Biggr] \; , \label{flat-potential}
\end{align}
 with $\varepsilon \ll 1$.  A similar setup was considered in the context of a Swampland conjecture in Ref.~\cite{Kadota:2019dol}. This is what we have in our mind implicitly, but in the actual numerical analyses, we turn off $\varepsilon $ and just discuss to what extent the field value can be extended without violating the Swampland conjectures.  
 
The above potential in the limit $\varepsilon \to 0$ is the T-model realization of the inflationary $\alpha$-attractor~\cite{Ellis:2013nxa, Ferrara:2013rsa, Ferrara:2013eqa, Ferrara:2014rya, Kallosh:2013yoa, Kallosh:2014rga, Galante:2014ifa, Cecotti:2014ipa, Carrasco:2015pla, Roest:2015qya, Linde:2015uga, Scalisi:2015qga}. This can be seen as a generalization of the $R^2$ model~\cite{Starobinsky:1980te} and the Higgs inflation model~\cite{Bezrukov:2007ep}.  It is obtained by canonical normalization of the scalar field $\varphi$ in the model's defining frame with the kinetic term  $-\frac{1}{2} \left(1 - \frac{\varphi^2}{6 \alpha} \right)^{-2} (\partial_\mu \varphi)^2$ and a generic scalar potential containing the mass term $- \frac{1}{2} m^2 \varphi^2$.\footnote{
Flattening of the potential also happens in the absence of the pole~\cite{Takahashi:2010ky, Nakayama:2010kt}.  Cases with different orders of the pole are studied in Refs.~\cite{Galante:2014ifa, Broy:2015qna, Terada:2016nqg}.
}  $\alpha$ is a dimensionless parameter of the model, and the overall scale $V_0$ is related to the mass as $V_0 = 3 \alpha m^2$.   $\phi$ has an approximate shift symmetry, and the small $\varepsilon$ term breaks it, as explained above.  
Qualitative features of our results below do not depend on the details of the model.
An advantage of the above setup is that the model becomes a large-field model as well as a small-field model depending on the value of $\alpha$ since the typical field range scales as $\Delta \phi \sim \sqrt{6 \alpha}$, so we expect our analyses are not too specific. We have also studied the quadratic potential and quadratic/quartic hilltop potentials, but we have not found clear evidence of qualitatively different results.

\begin{figure}[t] 
\begin{center}
\includegraphics[width=0.55\columnwidth]{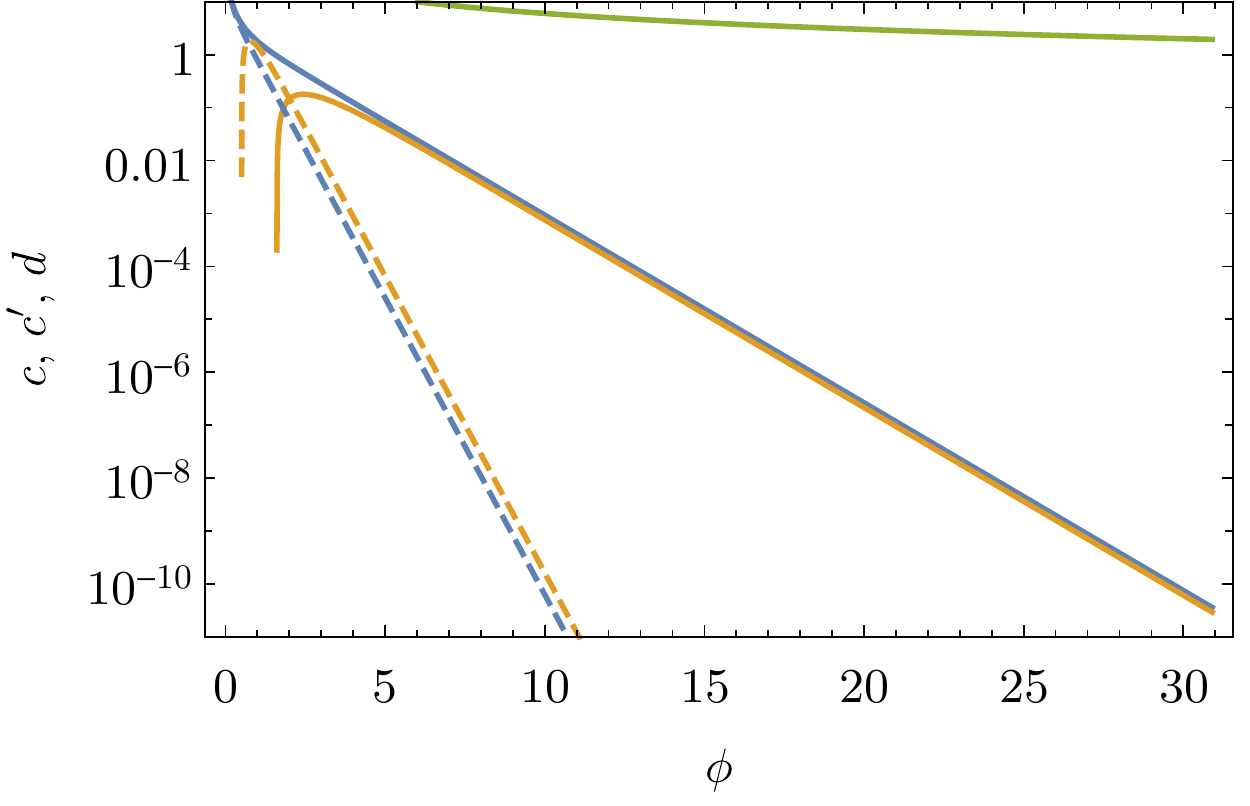}
\caption{ The maximum values of the parameters $c$ (blue), $c'$ (orange, decreasing in the left side), and $d$ (green, the top) in the Swampland inequalities for the potential~\eqref{flat-potential}.  The solid and dashed lines correspond to $\alpha = 1$ and $0.1$, respectively. The upper bound on $d$ does not depend on $\alpha$, and those on $c$ and $c'$ do not depend on $V_0$, which is set to be $2.0\times 10^{-9}$.  
}
\label{fig:swampland-parameters}
\end{center}
\end{figure}

Later, we will discuss the field excursion in the large $\phi/\sqrt{\alpha}$ region where the resulting e-folding number is large.  In this region, the slow-roll parameters are small.  If these are compatible with the Swampland conjectures, the parameters in the conjectures are bounded from above.  Such maximum values in terms of the field value $\phi$ are shown in Fig.~\ref{fig:swampland-parameters}.  In the large $\phi/\sqrt{\alpha}$ region, the upper bounds are approximated as $c < 4 \sqrt{2/(3 \alpha)}  \exp(- \sqrt{2/(3\alpha)} \phi) $, $c' < (8/(3\alpha)) \exp (-\sqrt{2/(3\alpha)}\phi) $, and $d < (3/\phi) \log (M_\text{P}/H) $~\cite{Scalisi:2018eaz}.

We now discuss the magnitude of the overall scale of the potential $V_0$.  In principle, the inflation just after the nucleation of the universe needs not to be responsible for the cosmic microwave background (CMB) fluctuations that we observe, and there may be second (or later) inflation.  The energy scale of the first inflation must be at least higher than that of the observable inflation scale, i.e., $V_0 \geq V_\text{CMB}$.  From the non-detection of the tensor mode in the CMB observations and from the requirement of successful big-bang nucleosynthesis, the energy scale of the observable inflation $V_\text{CMB}$ is restricted as $4.0 \times 10^{-76} (=(3.4 \times 10^{-3} \, \text{GeV})^4)  \leq  V_\text{CMB} \leq 2.0 \times 10^{-9} (=(1.6 \times 10^{16} \, \text{GeV})^4)$ where we used a lower bound on the reheating temperature $T_\text{R} \gtrsim 4 \, \text{MeV}$~\cite{Hannestad:2004px, deSalas:2015glj, Hasegawa:2019jsa},  the upper bound on the tensor-to-scalar ratio $r < 0.065$~\cite{Akrami:2018odb}, and the normalization of the scalar perturbations, $A_\text{s} = 2.1 \times 10^{-9}$~\cite{Akrami:2018odb}.  Because of the probability weight $\exp(-24\pi^2/V)$, however, creation of the universe with smaller $V_0$ is exponentially favored.  This implies that a natural choice of $V_0$ would be only slightly above $V_\text{CMB}$, so $V_0 \lesssim 2.0 \times 10^{-9} = (1.6 \times 10^{16} \, \text{GeV} )^4$.  
That is, even if there are several regions in the field space that allows inflation with various energy scales, the probabilities associated with those with $V \gg V_\text{CMB}$ are suppressed exponentially. 

An exception to the above discussion can arise when extremely long inflation such as eternal inflation is possible with higher $V_0$.  In this case, the volume-weight factor in eq.~\eqref{volume-weighted_probability} enhances the probability of the creation of the universe with a higher value of the potential.  This requires an additional assumption that the universe created in this way can successfully reach the part of the scalar-field manifold where the inflation relevant for the CMB observation occurs.  Thus, $V_0$ can be in principle higher than $V_\text{CMB}$ and treated as a free parameter.  However, since this possibility depends on the details of the landscape of the multifield potential, we use $V_0 = 2.0 \times 10^{-9} $ as a benchmark value.

\subsection{Analytic estimates} \label{sec:analytic_estimates}

In the following, we point out a simple fact that the volume factor solution to the small probability issue of the no-boundary proposal is severely constrained by the Swampland conjectures. As discussed around eq.~\eqref{minimum-probability}, the volume factor becomes relevant when the eternal inflation takes place.  The latter is severely constrained, if not excluded, by the Swampland conjectures~\cite{Matsui:2018bsy,Dimopoulos:2018upl,Kinney:2018kew,
Brahma:2019iyy}.\footnote{
On the connection between the eternal inflation and Swampland conjectures, see also Refs.~\cite{Rudelius:2019cfh, Wang:2019eym, Blanco-Pillado:2019tdf}.
}
   This seems clear, but let us discuss it more quantitatively.  We present a rough argument here and demonstrate it numerically in the next subsection.  

The necessary volume factor for $P_\text{inf}$ to overcome $P_\text{non-inf}$ is obtained from the condition $3 \mathcal{N}_e \simeq 24 \pi^2/V$.  For the dS conjecture to be still satisfied at this point, 
\begin{align}
c <&  \sqrt{ \frac{3 \alpha }{2}} \frac{V_0}{8 \pi^2} \; ,    &     \text{or} &    &
c' < &  \frac{V_0}{8 \pi^2} \; ,   \label{constraints_c}
\end{align}
must be satisfied 
  where we have used $\epsilon = 3 \alpha / (4 \mathcal{N}_e^2)$ and $\eta = - 1/ \mathcal{N}_e$ in the above model.  One can straightforwardly generalize this analysis to any inflationary models whose slow-roll parameters can be parametrized by a function of the e-folding number.  Several members of the universality classes of inflation~\cite{Mukhanov:2013tua, Roest:2013fha, Garcia-Bellido:2014gna, Binetruy:2014zya} (including the monomial chaotic, hilltop and inverse-hilltop models) are characterized as $\epsilon \propto 1/ \mathcal{N}_e^k$ with $k \geq 1$.  In the least constrained case ($k=1$; the monomial chaotic models), the maximally allowed $c$ scales as the square root of the potential value evaluated at the South Pole.

When the condition $3 \mathcal{N}_e \simeq 24 \pi^2/V$ is satisfied, the field value is given by $\phi \simeq \sqrt{\frac{3 \alpha} {2}} \log  \frac{ 64 \pi^2 }{3 V_0}$.  For the distance conjecture to be still valid at this point, 
\begin{align}
 d \lesssim \sqrt{\frac{3}{ 2 \alpha} } \; ,     \label{constraint_d}
\end{align}
must be satisfied. Even if the refined dS conjecture is omitted, this gives a nontrivial constraint for $\alpha \gtrsim \mathcal{O}(1)$.

As we have briefly discussed in Sec.~\ref{subsec:Hartle-Hawking-Hertog}, the extrapolation of the slow-roll formulae into the eternal inflation region is not justified.  A more conservative estimate on the field value corresponding to the required volume-weight factor is obtained by simply requiring that the eternal inflation happens (see the condition~\eqref{eternity_condition}).   This leads to~\cite{Matsui:2018bsy,Dimopoulos:2018upl,Kinney:2018kew,
Brahma:2019iyy}
\begin{align}
c <& \frac{1}{2\pi} \sqrt{\frac{V_0}{3}} \; ,   &     \text{or} &    &
c' < 3 \;. \label{conservative_constraints_c}
\end{align}
It is easy to satisfy the second inequality, but to realize eternal inflation, the first derivative of the potential must be suppressed.  Once one specifies the potential, the first and second derivatives usually correlate with each other.  For example, with our example potential~\eqref{flat-potential}, it is insufficient to require $c'<3$, and the compatibility with eternal inflation actually implies $c < \frac{1}{2\pi} \sqrt{\frac{V_0}{3}} $ or $c' < \frac{1}{3\pi} \sqrt{\frac{V_0}{2 \alpha }} $.
Similarly, a conservative version of the constraint on the parameter of the distance conjecture is
\begin{align}
d \lesssim \sqrt{\frac{6}{\alpha}} \; .     \label{conservative_constraint_d}
\end{align}
The constraints~\eqref{conservative_constraints_c} are parametrically milder than the naive constraint~\eqref{constraints_c}, while the constraint~\eqref{conservative_constraint_d} is similar to the naive one~\eqref{constraint_d}.

It is remarkable that even a fairly conservative assumption of sub-Planckian energy $V_0^{1/4} \lesssim 0.1 M_\text{P}$ combined with the codition~\eqref{conservative_constraints_c} leads to nontrivial inequalities, $c, \, c'\sqrt{\alpha} \lesssim \mathcal{O}(10^{-3})$, which are stronger than 
the CMB bound~\cite{Agrawal:2018own, Achucarro:2018vey, Garg:2018reu, Kinney:2018nny, Brahma:2018hrd, Das:2018hqy, Fukuda:2018haz, Ashoorioon:2018sqb}.

\subsection{Numerical analyses}

We apply the formalism reviewed in Section~\ref{sec:no-boundary} to the model~\eqref{flat-potential} under consideration. The approximate expressions~\eqref{pprobability} and \eqref{volume-weighted_probability} of the probability are not precise enough around the point where inflation ends since the slow-roll condition ceases to be a good approximation.  Numerical evaluations are required. 

In our numerical evaluation, the unit is chosen in such a way that $1/\ell^2 = V(\phi_\text{SP}^\text{R})$ for each choice of $\phi_\text{SP}^\text{R}$.  In this convention, there are no hierarchically large numbers when solving the equations of motion.   When the slow-roll approximation holds, $a$ and $\phi$ becomes real on the line $\tau^\text{R} \simeq \sqrt{3} \pi /2$ for sufficiently large $\tau^\text{I}$ (see also Appendix~\ref{sec:generic_slow-roll}).  

\begin{figure}[thb!] 
\begin{center}
   \subcaptionbox{$\ln |a^\text{I}|$ \quad  (quasi-dS) \label{sfig:a_quasi-dS}}
{\includegraphics[width=0.48\columnwidth]{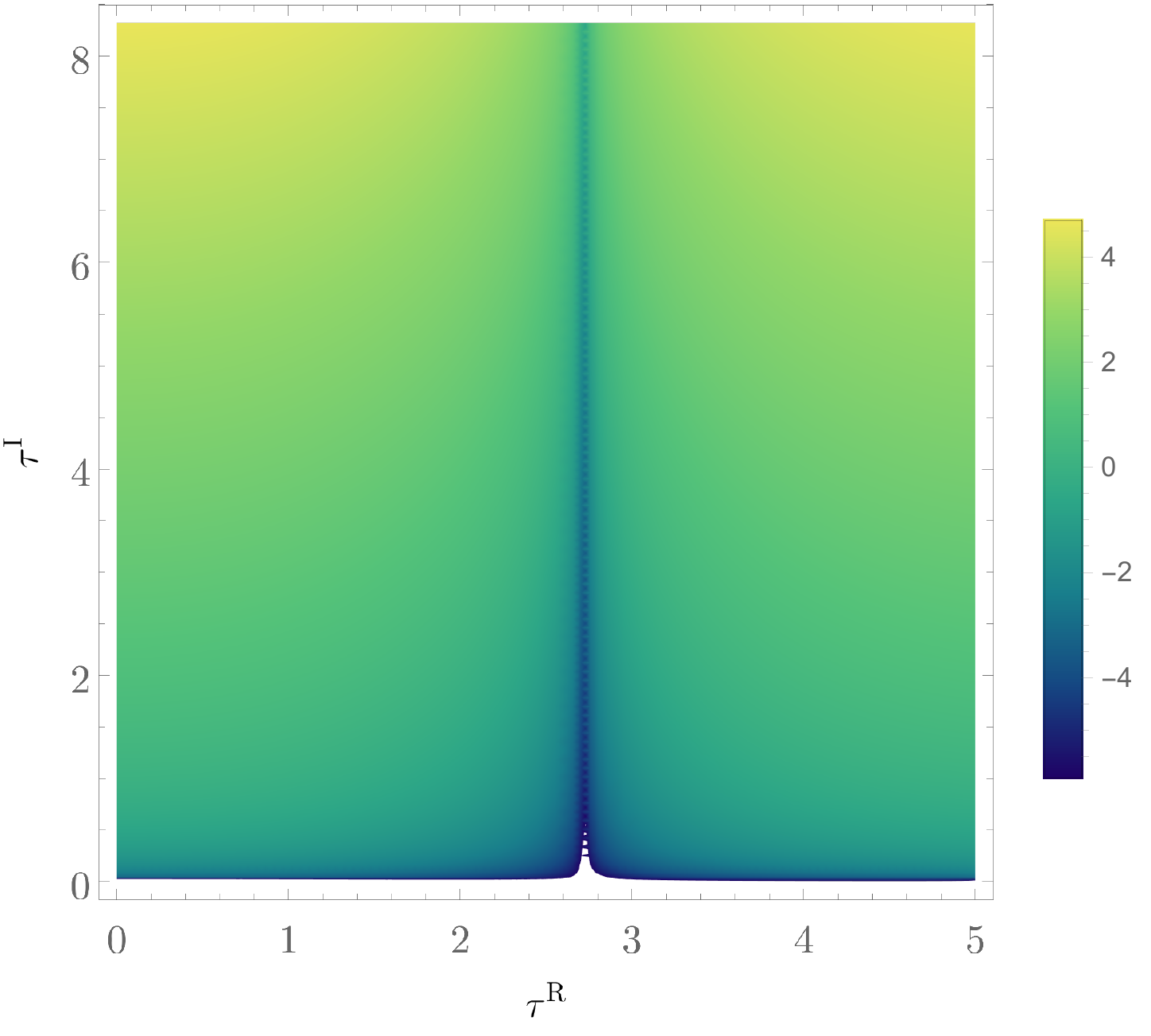}}~~~~~~~~~
   \subcaptionbox{$\ln |a^\text{I}|$ \quad  (non-dS) \label{sfig:a_non-dS}}
{\includegraphics[width=0.48\columnwidth]{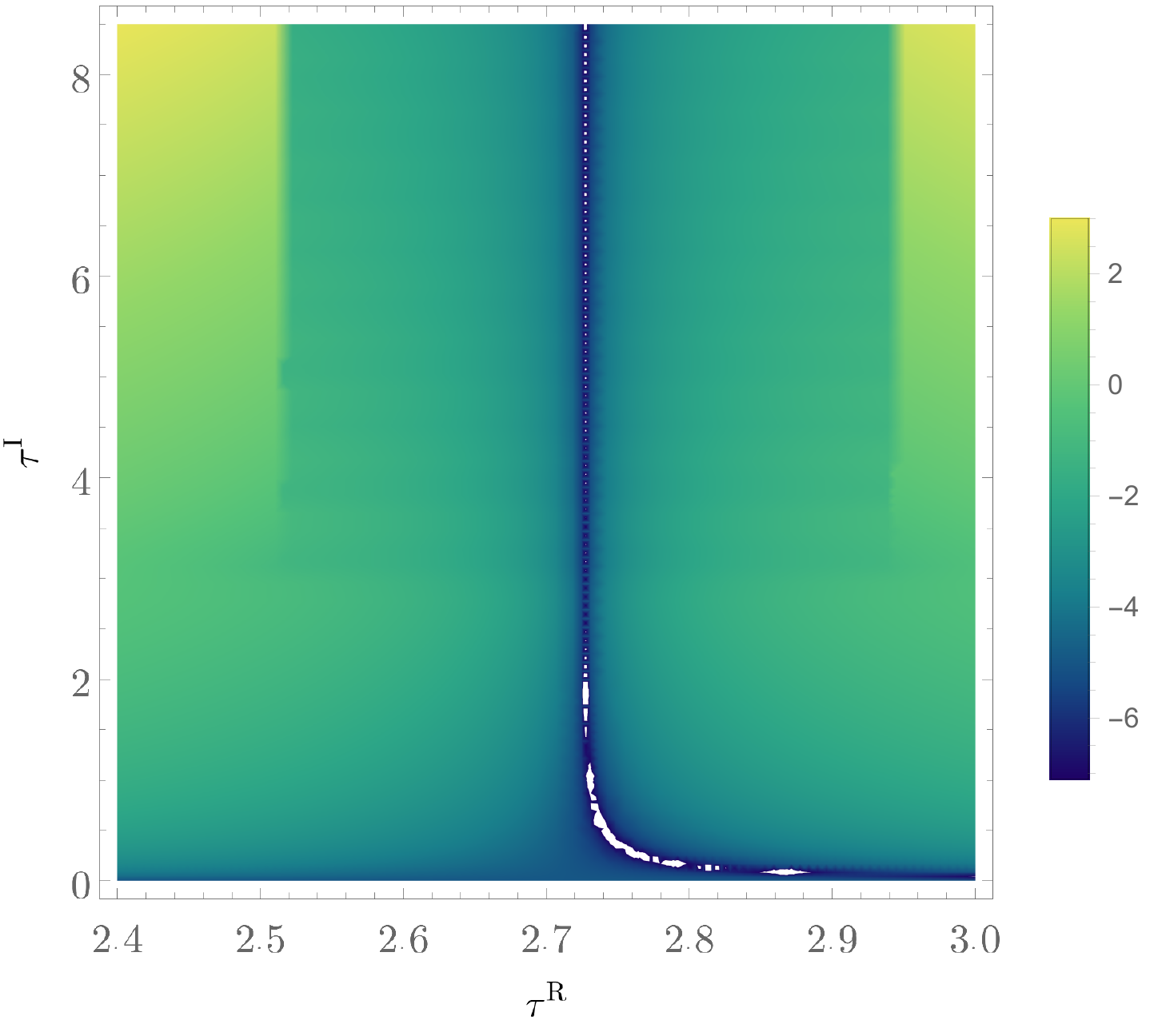}} \\ 
\vspace{2mm}
   \subcaptionbox{$\ln |\phi^\text{I}|$ \quad  (quasi-dS) \label{sfig:phi_quasi-dS}}
{\includegraphics[width=0.48\columnwidth]{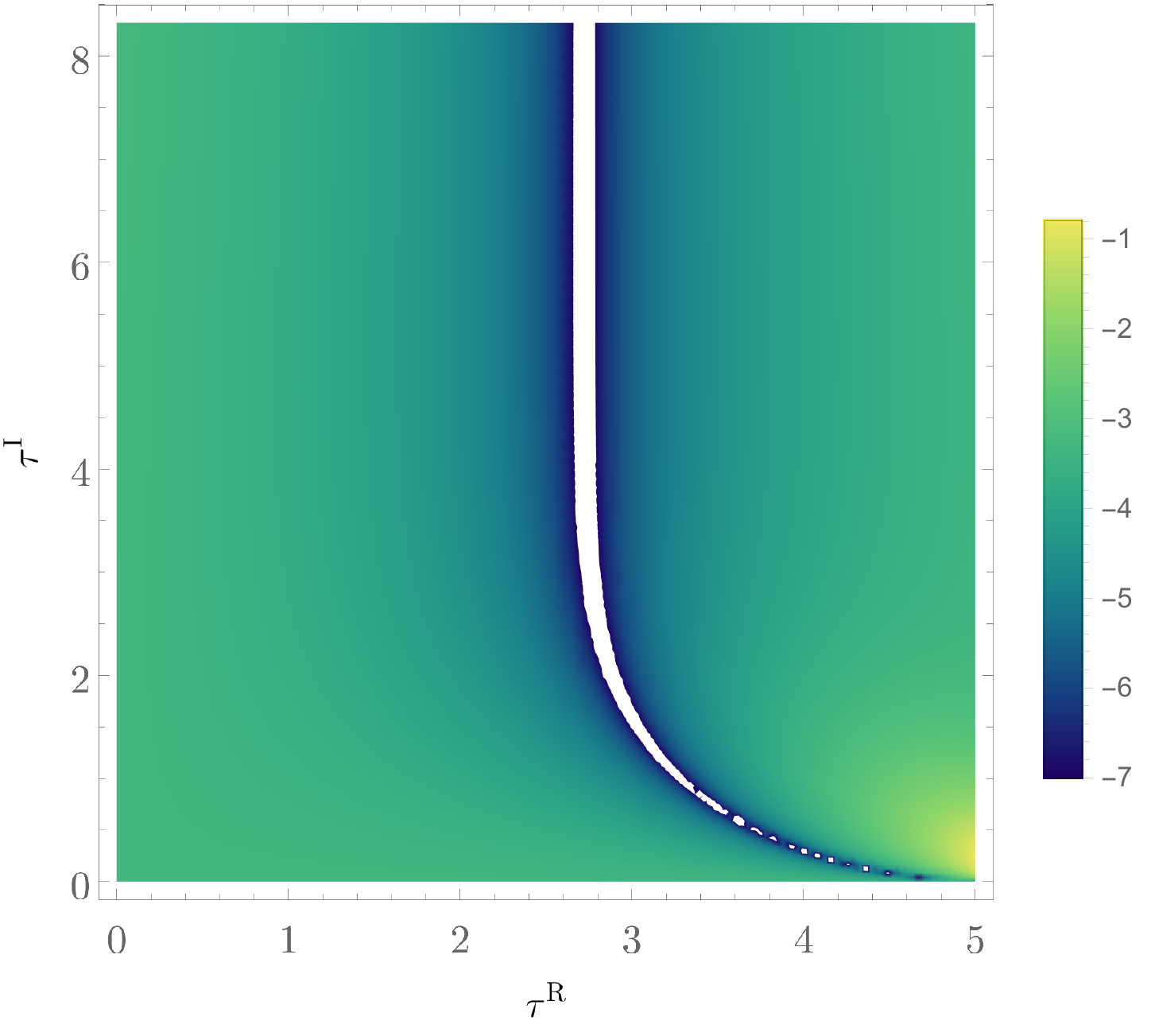}}~~~~~~~~
   \subcaptionbox{$\ln |\phi^\text{I}|$ \quad  (non-dS) \label{sfig:phi_non-dS}}
{\includegraphics[width=0.48\columnwidth]{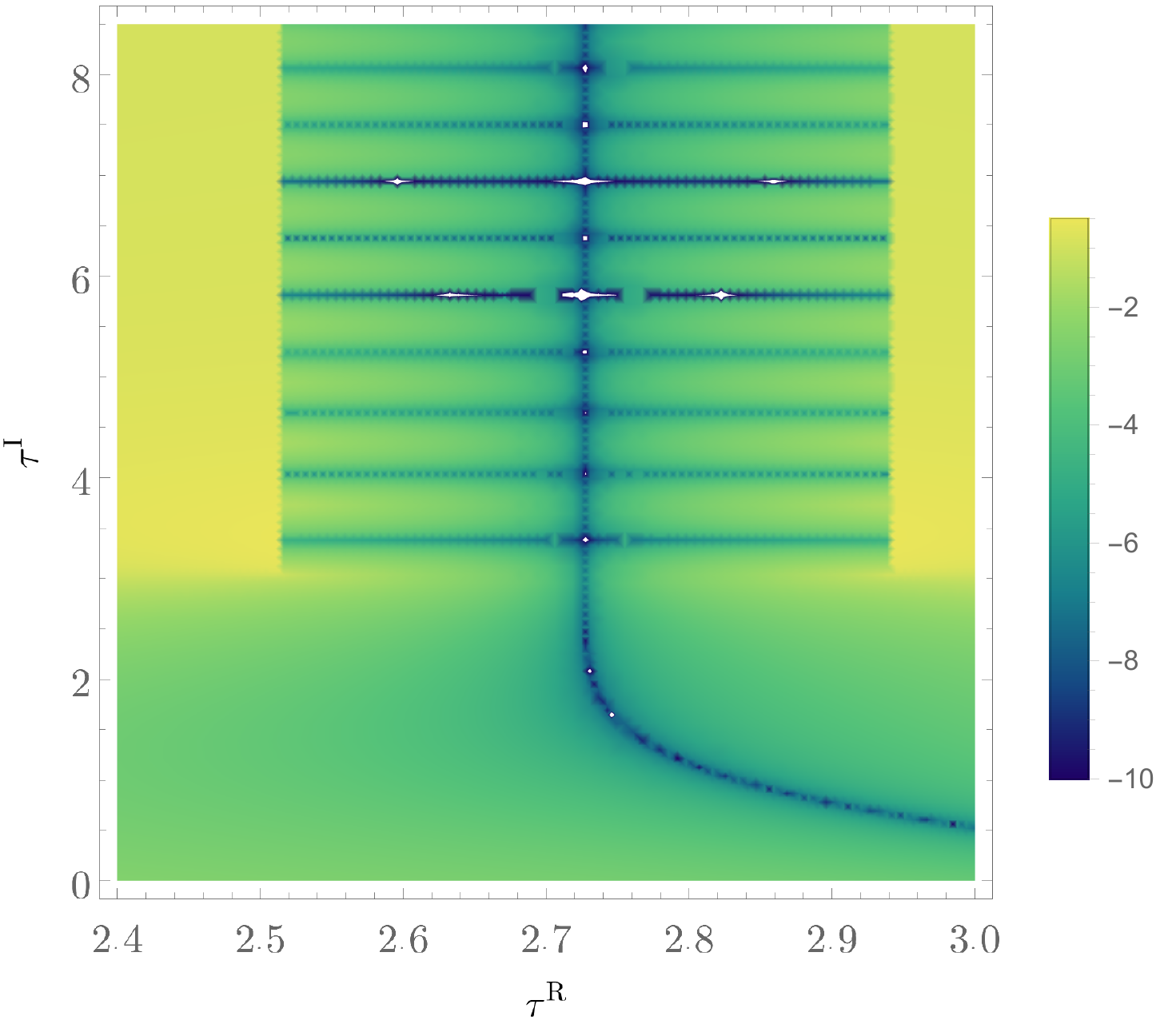}}
\caption{ Density plots of the imaginary part of $a$ (top) and of $\phi$ (bottom) on the complex $\tau$ plane. The color represents $\ln |a^\text{I}|$ and $\ln |\phi^\text{I}|$.  (Left) The parameters are $\alpha = 1$ and $\phi_\text{SP} = 6.06127 - (3.61564 \times 10^{-2}) i$, representing a quasi-dS regime.  (Right) The parameters are $\alpha = 0.01$ and $\phi_\text{SP} = 0.700503 - 0.154991 i$, representing a non-dS regime.  The part of the nontrivial structure is focused, so the region around the origin is not shown.  
}
\label{fig:density-plot}
\end{center}
\end{figure}

The probability distribution with respect to $\chi$ in eq.~\eqref{probability} is (by definition) independent of $b$ (though small $b$ dependence arises from approximations, as mentioned above). We set $b = \sqrt{3} \exp{2.5}$,\footnote{
The field-dependent normalization convention ($1/\ell^2 = V(\phi_\text{SP}^\text{R})$) effectively corresponds to a $\chi$-dependent choice of $b$ in a field-independent normalization convention.  This affects the proper definition of the probability density~\eqref{probability}, but the change is subdominant compared to the exponential factor $\exp (- 2 S_\text{E}^\text{R})$, so we neglect it.
} which roughly means that the universe expands 2.5 e-folding in the Lorentz but still quantum regime. The specific value 2.5 is not important, but if it is much smaller, there are no solutions that allow simultaneously real $b$ and $\chi$.  This can be seen clearly in Figs.~\ref{sfig:a_quasi-dS} and \ref{sfig:phi_quasi-dS}, which show the imaginary part of $a$ and $\phi$ on the complex $\tau$ plane.  There are no overlaps between zeros of $a^\text{I}$ and of $\phi^\text{I}$ in the region with small $\tau^\text{I}$.  On the other hand, if it is much larger, numerical errors accumulate, so the classicality condition cannot be confirmed.  This technical issue has already been reported in Ref.~\cite{Battarra:2014kga}. 

There are no singularities (poles nor branch points/cuts) in Figs.~\ref{sfig:a_quasi-dS} and \ref{sfig:phi_quasi-dS}, but we found branch points/cuts for small $\alpha$ when the slow-roll condition is not well satisfied.  The density plot for such a case is shown in Figs.~\ref{sfig:a_non-dS} and \ref{sfig:phi_non-dS}. One can see branch points at $\tau \simeq 2.52 + 3 i $ and $\tau \simeq 2.94 + 3 i$ and brach cuts emanating from them.  The direction of branch cuts is in the direction in which we solve the equations of motion numerically.  The structure like ``fishbone'' represents the way $\phi$ oscillates around the minimum of the potential.  We find that the position and the number of branch points are model dependent. 
For smaller values of $\phi_\text{SP}^\text{R}$, we can no longer find solutions which satisfies the boundary conditions $a^\text{R} = b$ and $a^\text{I} = \phi^\text{I} = 0$.  This is consistent with the statement in the literature~\cite{Hartle:2007gi,Hartle:2008ng} that the classical solution disappears when $\phi$ becomes small.

To avoid the branch cuts in our main analyses, we solve the equations of motion in the Euclidean (real $\tau$) direction from the origin $\tau = 0$ to a point $\tau = t_0(>0)$.  We then solve them in the Lorentzian direction to reach $\tau = t_0 + i t_1$ where $t_1>0$.  We again solve them in the Euclidean direction to reach $\tau = t_2 + i t_1$ where $t_2 > t_0$.  As long as the branch points/cuts are avoided, the values of these parameters, $t_0, t_1$, and $t_2$, are unimportant. 
  In the last segment of the Euclidean part, we numerically solve the time at which $a^\text{I}$ and $\phi^\text{I}$ vanishes.  Then, we change $\phi_\text{SP}^\text{I}$ so that these time approaches to each other.   Finally, we solve the equations of motion in the Lorentzian direction again to find the value of $\tau^\text{I}$ that satisfy $a^\text{R} = b$.  We iterate this procedure until the errors get within the specified precision ($10^{-8}$).   In this way, we obtain $\phi_\text{SP}^\text{I}$ and also $\chi$ from the input values $(b, \phi_\text{SP}^\text{R})$.  This is equivalent to say that $\phi_\text{SP}$ is determined from $b$ and $\chi$ (up to an uncertainty discussed in the next paragraph).

\begin{figure}[t] 
\begin{center}
   \subcaptionbox{$\alpha = 1$ \label{sfig:phi-chi1}}
{\includegraphics[width=0.46\columnwidth]{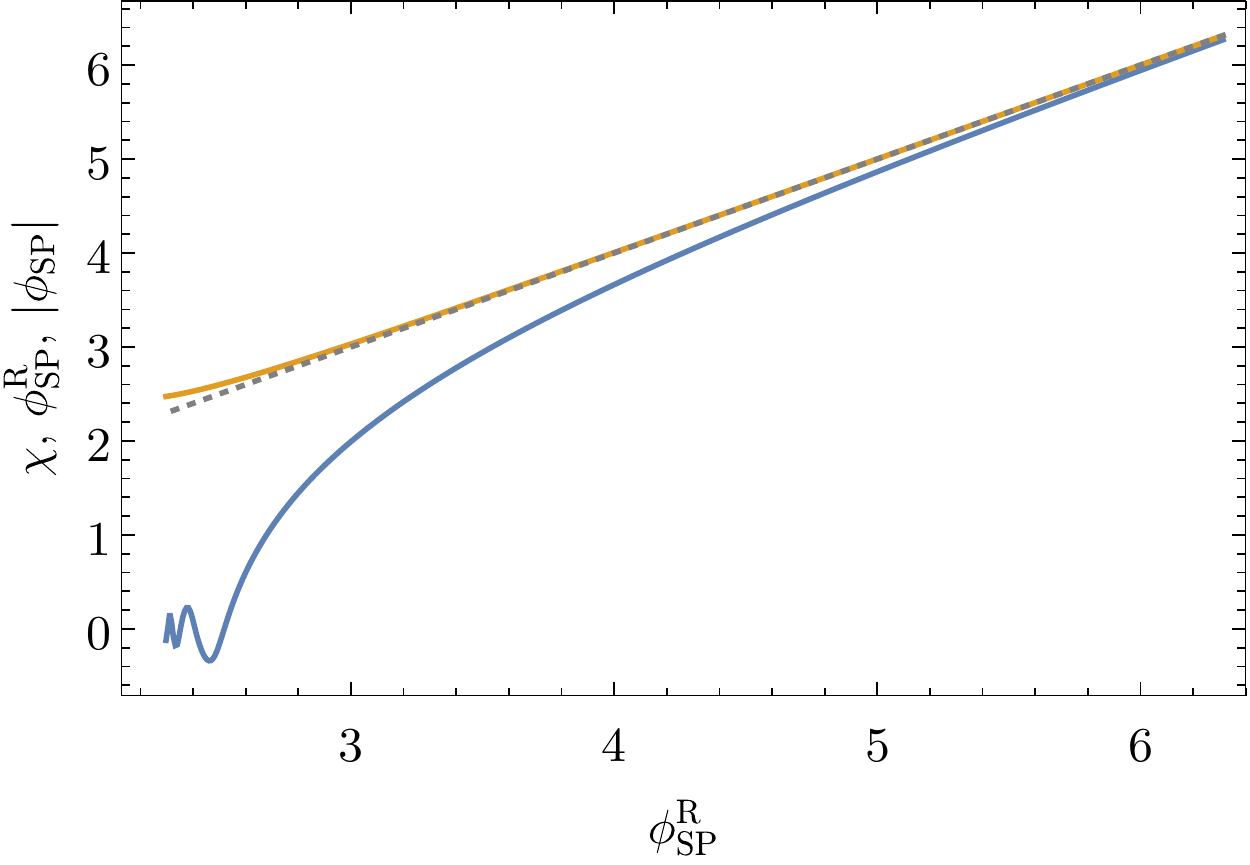}}~~~~~~
   \subcaptionbox{$\alpha = 0.1$\label{sfig:phi-chi01}}
{\includegraphics[width=0.47\columnwidth]{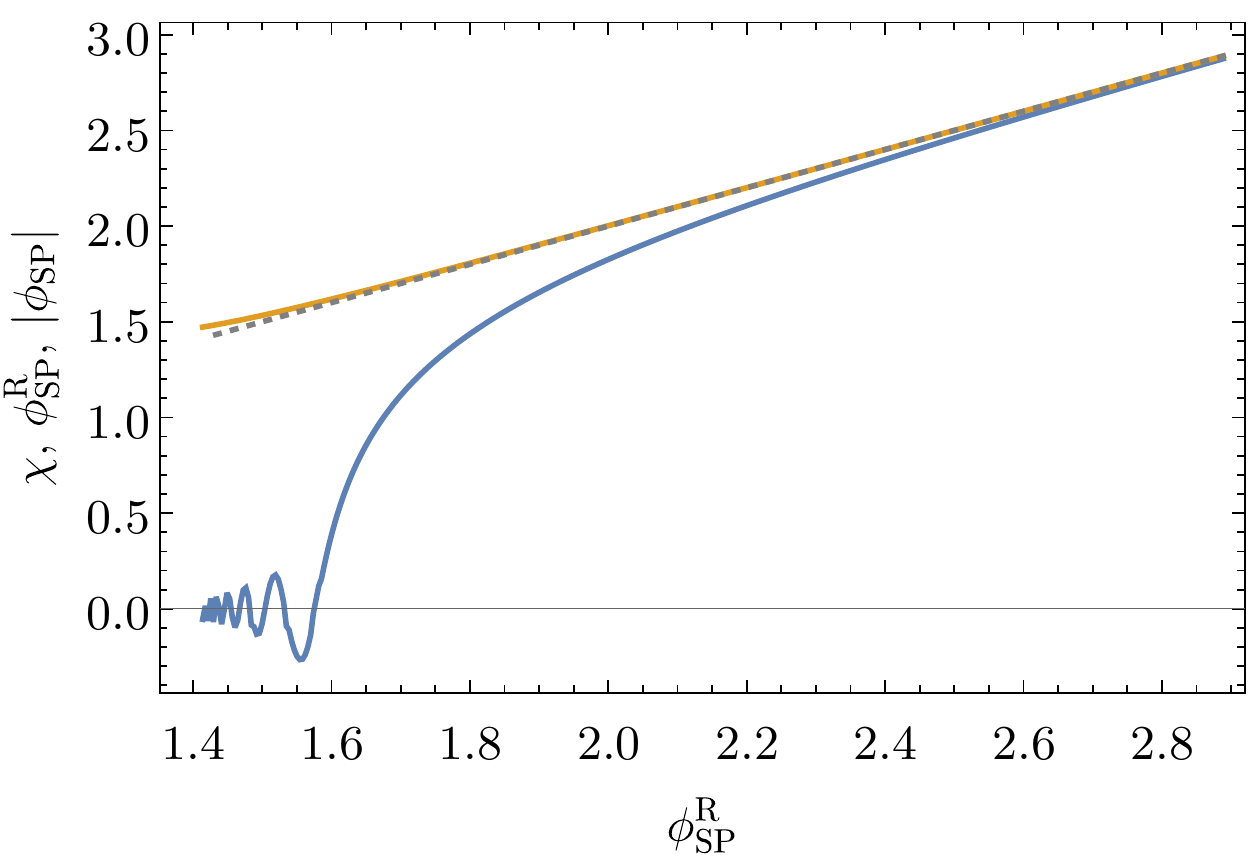}} 
\caption{The relation between the scalar field values at the South Pole ($\tau = 0$ at which $a=0$ and $\phi =\phi_\text{SP}$) and at the nucleation point (at which $\phi = \chi$ and $a = b$).  The blue solid (bottom) lines show $\chi$, and the orange solid (top) lines show $|\phi_\text{SP}|$. To guide the eyes, the line of $\phi_\text{SP}^\text{R}$ (same as the horizontal axis) is shown as the gray dotted lines.  The value of $b$ is taken as $b = \sqrt{3} \exp (2.5)$.   }
\end{center}
\end{figure}

The relation between the South Pole scalar field value $\phi_\text{SP}$ and the value at nucleation $\chi$ is plotted in Figs.~\ref{sfig:phi-chi1} and \ref{sfig:phi-chi01}.  When the field value is large so that the slow-roll conditions are well satisfied, the difference between $\phi_\text{SP}^\text{R}$ and $\chi$ is small. In this regime, the relation between $\phi_\text{SP}^\text{R}$ and $\chi$ as well as $\phi_\text{SP}^\text{R}$ and $\phi_\text{SP}^\text{I}$ is monotonic.   On the other hand, the difference becomes substantial in the opposite limit, and the monotonicity is lost as $\chi$ oscillates around the minimum. 
Because of the oscillating feature, we use $\phi_\text{SP}^\text{R}$ as an input variable in some of the following figures although $\chi$ is a more fundamental variable.

\begin{figure}[thb!] 
\begin{center}
\includegraphics[width=0.47\columnwidth]{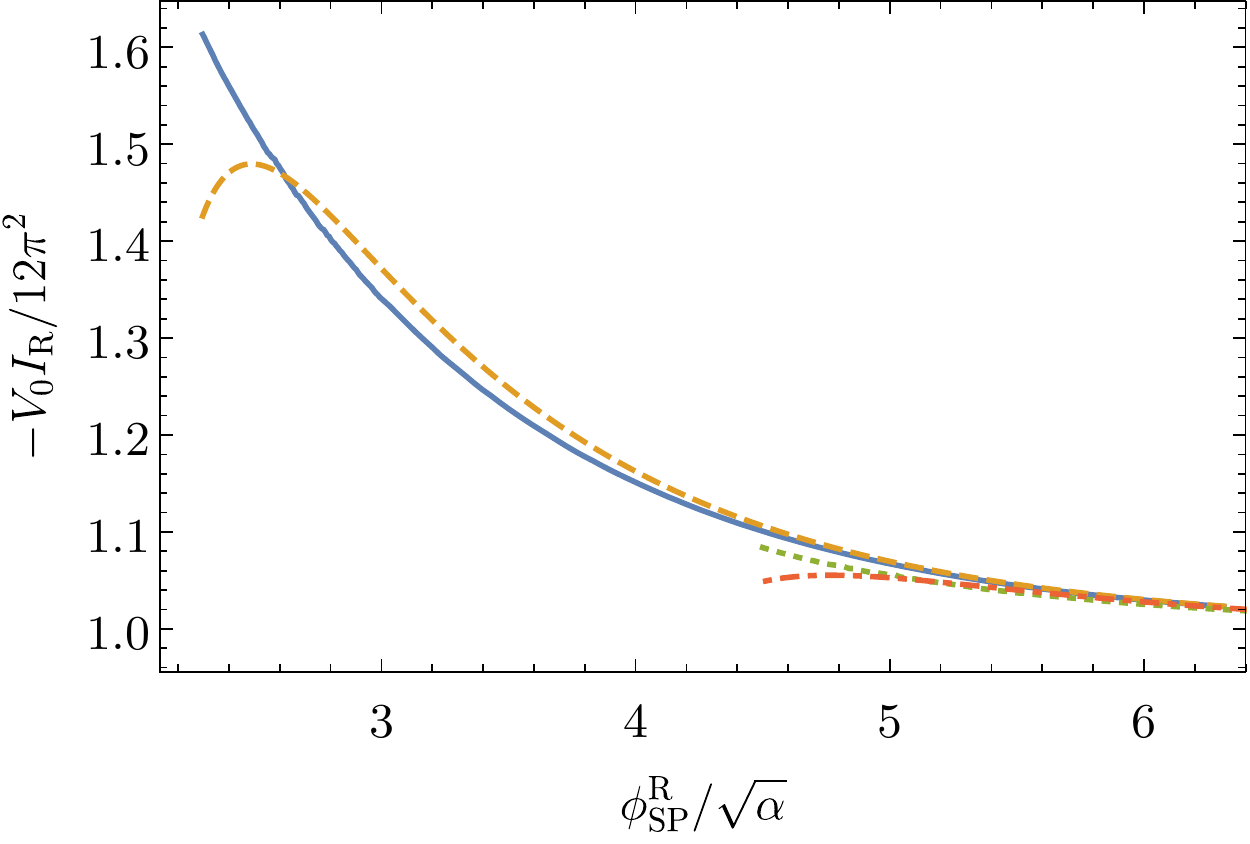}~~~~~~
\includegraphics[width=0.47\columnwidth]{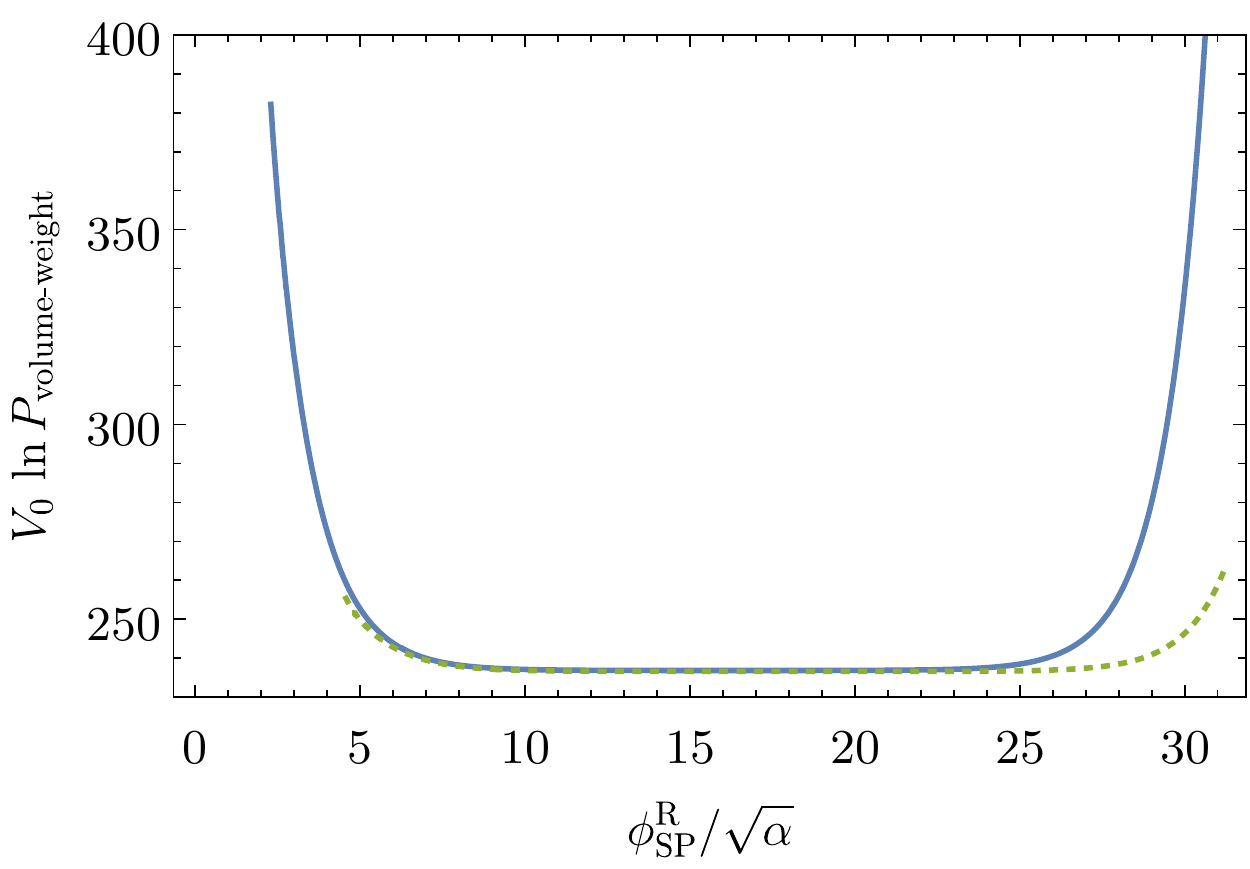}
\caption{(Left) The real part of the Euclidean action as a function of $\phi_\text{SP}^\text{R}$ in units of $\sqrt{\alpha}$. The normalization is such that it asymptotes to unity in the dS limit.  The blue solid and orange dashed lines denote the numerical and approximate values, respectively, for $\alpha =1$.  The shorter, green dotted and red dot-dashed lines denote the numerical and approximate values, respectively, for $\alpha = 0.1$.  (Right) With the volume factor (obtained from the slow-roll formula).  $V_0 = 2.0 \times 10^{-9}$ is assumed.   The blue solid and green dotted lines denote $\alpha = 1$  and $0.1$, respectively, corresponding to the same-color lines in the left panel.  The numerical results are used on the left side of the lines, and they are extended to the right side by the approximate analytic formulae. 
}
\label{fig:probability}
\end{center}
\end{figure}

The logarithm of the probability, i.e., the real part of the saddle-point value of the Euclidean action is shown in Fig.~\ref{fig:probability}.  The blue solid and green dotted lines denote the numerical result for $\alpha = 1$ and 0.1, respectively, while the orange dashed and red dot-dashed lines denote an approximate formula~\eqref{pprobability} for $\alpha = 1$ and 0.1, respectively.  The normalization in the left panel is such that it asymptotes to unity in the large-field slow-roll limit ($\phi_\text{SP}^\text{R} \to \infty$).  The plotted combination is independent of the overall scale of the potential $V_0$, but the actual probability sensitively depend on it: $\exp (- 2S_\text{E}^\text{R} ) \sim \exp (-  \text{Re} \, [24 \pi^2 / V(\phi_\text{SP})])$.  
 The numerical and approximate results agree with each other in the large-field region where the slow-roll condition is well satisfied.  Close to the left end of the lines, the difference  between the numerical and approximate results becomes non-negligible, and the approximate result gives a smaller value.  This is due to the growth of $\phi_\text{SP}^\text{I}$, which contributes to suppression of $\text{Re} \, V^{-1}(\phi_\text{SP})$.   This growth indicates the breakdown of the approximation used to obtain the approximate analytic result.  The slow-roll condition is still well satisfied at the South Pole, but that evaluated at the boundary hypersurface (where $\phi=\chi$ and $a=b$) is not.  For example, in the case of $\alpha =1$,  the slow-roll condition is violated around $\chi \simeq 1$, which corresponds to $\phi_\text{SP}^\text{R} \simeq 2.7$ (see Figs.~\ref{sfig:phi-chi1}).

No solution is found for $\phi_\text{SP}^\text{R} \lesssim 2.3$ (1.4) for $\alpha = 1$ (0.1) by our algorithm.  Although we cannot prove that there are indeed no solutions to satisfy the boundary condition, we assume their absence.  The disappearance of the classical solution is consistent with the literature~\cite{Hartle:2007gi,Hartle:2008ng} in which other potentials like the quadratic potential were studied.  
This implies that if we literally take $c = \mathcal{O}(1)$, no saddle-point solutions are allowed at all.

In the right panel of Fig.~\ref{fig:probability}, the conditional probability with the volume factor~\eqref{volume-weighted_probability} is shown assuming the slow-roll formula~\eqref{e-folding}.  Until the volume factor becomes relevant (so that the line goes up in the right side of the figure), this is the same as the unconditional probability (the left panel) up to the normalization.  For a sufficiently large field value, the e-folding number is so large that the volume factor outweighs the $1/V$ factor.  This needs longer field length for a smaller $V_0$ though the dependence is logarithmic.  For $\alpha =1$, the required length $\Delta \phi \sim 30$ is close to the maximal field distance allowed by the distance conjecture.  
In this figure, we compare the relative probability densities associated with the different parts of the same potential~\eqref{flat-potential}.  As we mentioned in Sec.~\ref{subsec:small-probability-issue}, however, a much larger e-folding number and corresponding larger field distance are necessary to exceed the probability of the direct creation of the universe at the current dark energy vacuum.  On the other hand, if we take the second (conservative) approach mentioned in Sec.~\ref{subsec:Hartle-Hawking-Hertog}, which requires only the occurrence of the eternal inflation, a much shorter field distance $\phi_\text{SP}^\text{R} / \sqrt{\alpha} \sim 17 $ (up to the logarithmic dependence on $\alpha V_0$) is enough to ensure that the probability of the classical inflationary universe is dominant.

\begin{figure}[thb!] 
\begin{center}
\includegraphics[width=0.47\columnwidth]{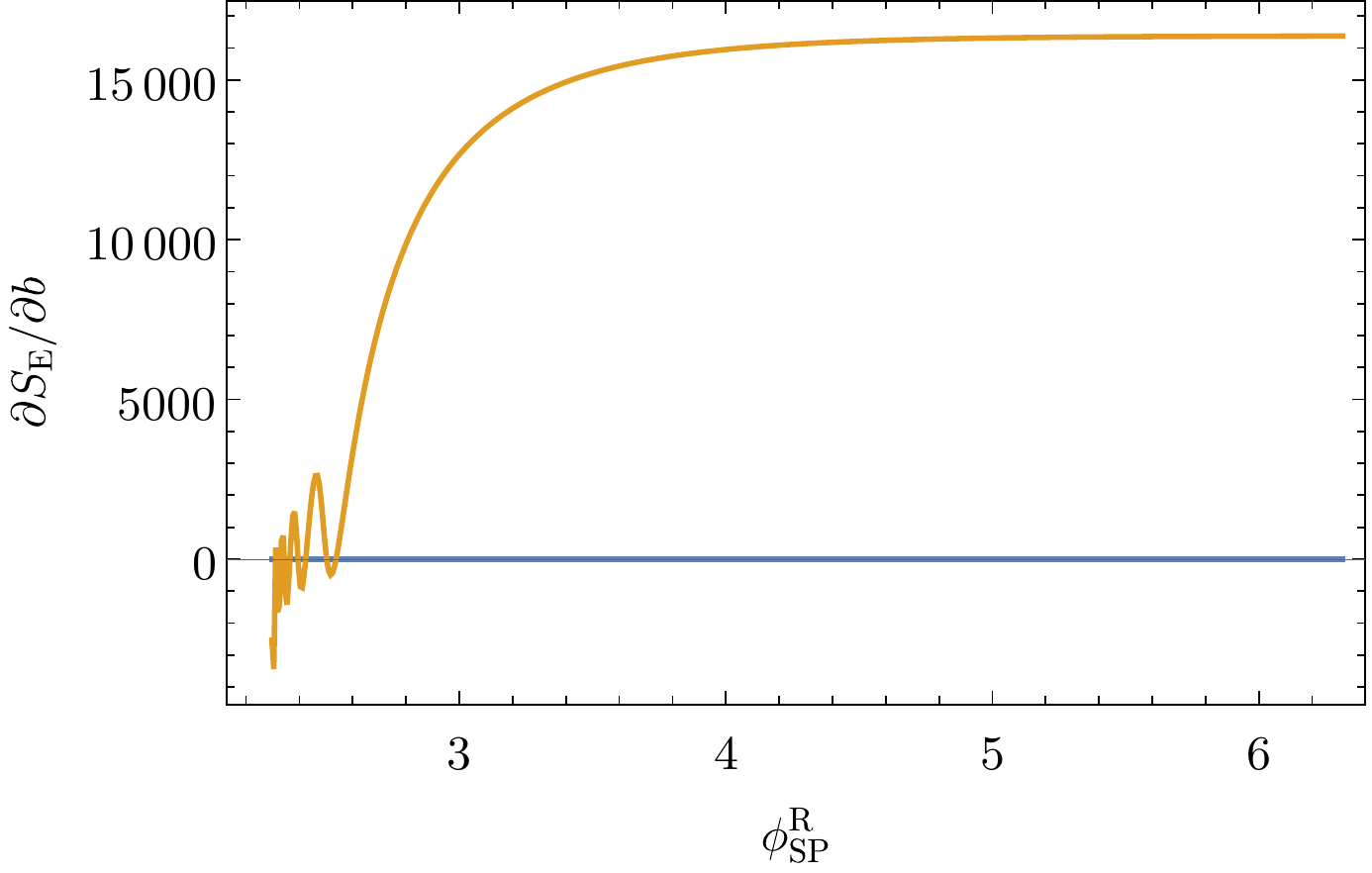}~~~~~
\includegraphics[width=0.47\columnwidth]{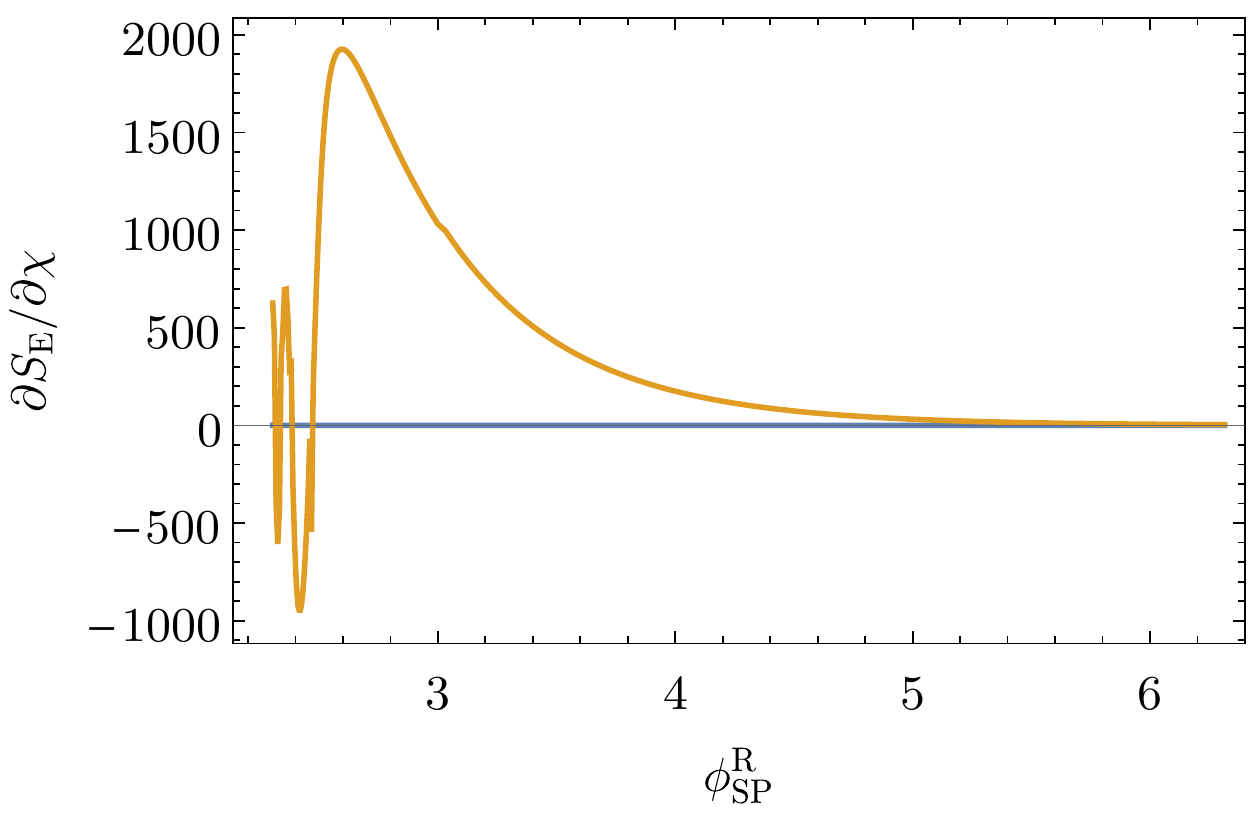}
\caption{The derivative of the Euclidean action.  The blue and orange lines correspond to the real and imaginary parts, respectively.  The classicality condition is satisfied in most parts. $\alpha = 1$ is assumed.
}
\label{fig:classicality}
\end{center}
\end{figure}

To check whether the above solution satisfies the classicality condition~\eqref{WKB}, we plot the derivative of the Euclidean action in Fig.~\ref{fig:classicality}.  The derivatives are approximately evaluated as the ratio of finite differences: $\partial S_\text{E} / \partial b \simeq \Delta S_\text{E} / \Delta b$ and likewise for the derivative with respect to $\chi$.  The blue and orange lines denote the real and imaginary parts, respectively. In most of the plotted region, the imaginary part is much larger than the real part, so the classicality condition is satisfied.  Because of the oscillating feature of $\chi$, the derivatives change their sign before the saddle-point solution disappears.  Thus, 
 the classicality is occasionally lost, and the saddle-point solution satisfying the boundary conditions finally disappears.  

\begin{figure}[t] 
\begin{center}
\includegraphics[width=0.47\columnwidth]{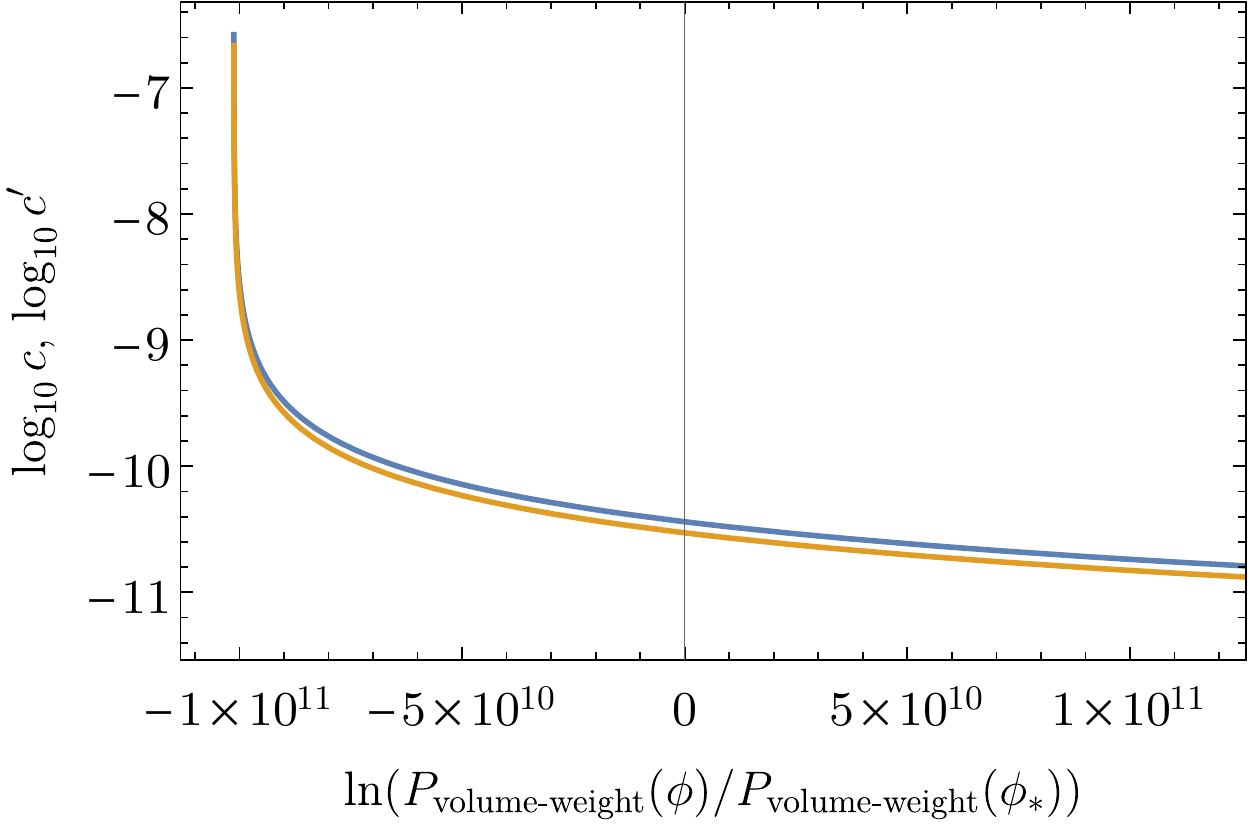}~~~~~
\includegraphics[width=0.47\columnwidth]{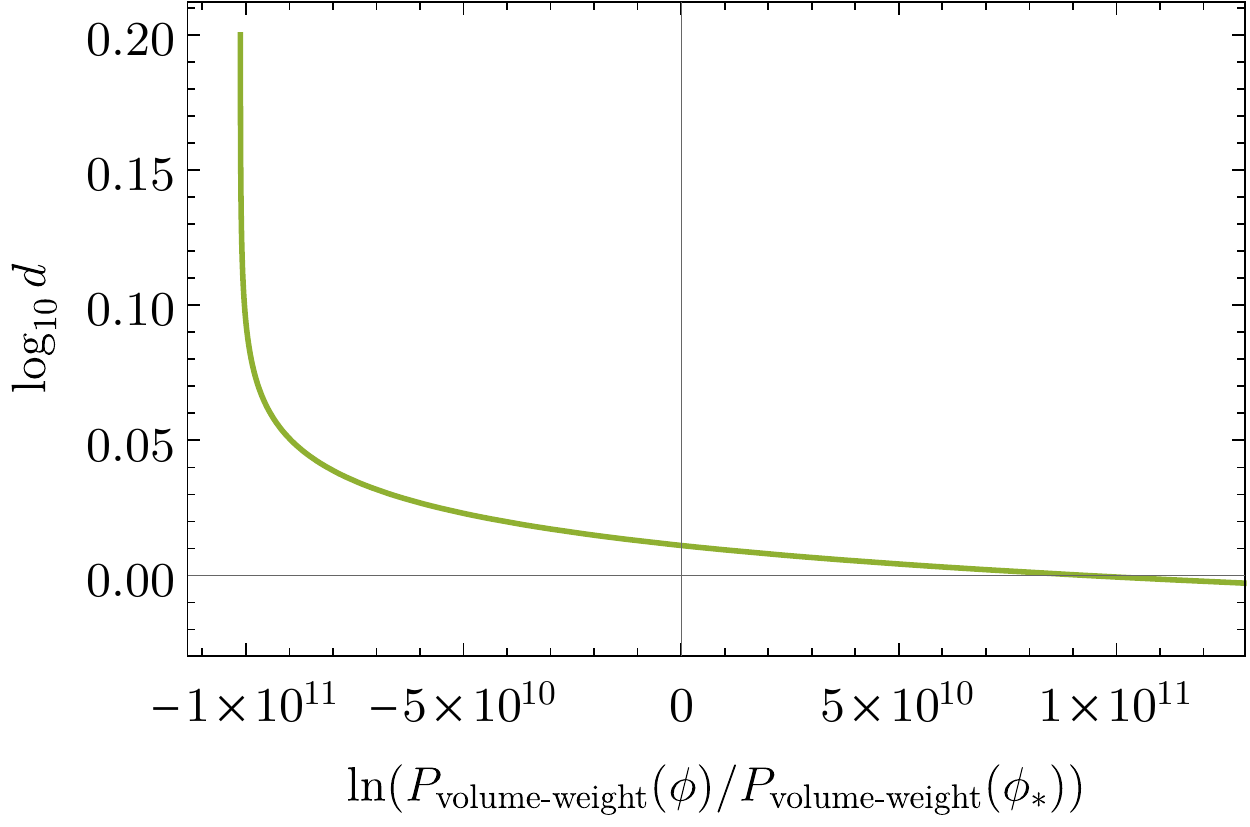}
\caption{The maximum values of the parameters in the Swampland conjectures as a function of (the logarithm of) the conditional probability density $P_\text{volume-weight}(\phi)$ at the point $\phi$ where the Swampland inequality is saturated, normalized by the probability density $P_\text{volume-weight}(\phi_*)$ at the point $\phi_*$ where the saddle-point solution disappears. In both panels,  $\alpha =1$ and $V_0 = 2.0 \times 10^{-9}$, and the slow-roll formula for the e-folding number is assumed.  (Left) The parameters of the (refined) dS conjecture $c$ (blue line; upper) and $c'$ (orange line; lower). (Right) The parameter of the distance conjecture $d$. 
}
\label{fig:constraints_on_SL-parameters}
\end{center}
\end{figure}

Finally, let us directly compare the parameters in the Swampland conjectures ($c$, $c'$, and $d$) and the logarithm of the probability evaluated at the point where the Swampland condition is saturated. This is shown in Fig.~\ref{fig:constraints_on_SL-parameters}.  Only the large-field region $\phi \gtrsim 20$ is plotted where the probability density $P(\phi)$ increases as $\phi$ increases (see also Fig.~\ref{fig:probability} (right); again, slow-roll formulae are extrapolated, corresponding to the constraints~\eqref{constraints_c} and \eqref{constraint_d}).      
Since the constraint on the parameters of the dS conjecture is much stronger than that on the parameter of the distance conjecture, the figures are separately plotted.  As anticipated in the previous subsection, for the volume factor solution and the dS conjecture to be compatible with each other, $c$ and $c'$ is constrained to be smaller than $\mathcal{O}(10^{-10})$.  The Figure shows the case of $\alpha = 1$ and $V_0 = 2.0 \times 10^{-9}$. (The dependence on these parameters $\alpha$ and $V_0$ has been discussed in Sec.~\ref{sec:analytic_estimates}.)  Similarly, the parameter of the distance conjecture $d$ should be $\mathcal{O}(1)$ or less.  Note also that the constraints on $c$ and $c'$ become much weaker if we only require the occurrence of eternal inflation rather than $P_\text{volume-weight}(\phi) > P_\text{volume-weight}(\phi_*)$ where $\phi_*$ is the value of $\phi$ where the saddle-point solution disappears.  For  $\alpha = 1$ and $V_0 = 2.0 \times 10^{-9}$, the inequalities~\eqref{conservative_constraints_c} and those below them tell us that they are smaller than $\mathcal{O}(10^{-5})$. 
Still, these are strong constraints on the  dimensionless parameters which are a priori expected to be $\mathcal{O}(1)$.

Even if these constraints on $c$, $c'$, or $d$ are only slightly violated, the probability is completely dominated by non-inflationary universe which will either collapse classically or stay in the quantum regime. This can be seen, e.g., from the scale of the horizontal axis in Fig.~\ref{fig:constraints_on_SL-parameters}, which is of order $-1 \times 10^{11}$.  The relative probability of the classical inflationary universe is roughly of order $\exp(-1 \times 10^{11}) \approx 10^{-5 \times 10^{10}}$ for $V_0 = 2.0 \times 10^{-9}$.    Of course, this sensitively depends on the value of $V_0$ as we repeatedly emphasized the dependence $P (\chi) \approx \exp (24 \pi^2 /V)$.  If we take, e.g., $c$, $c'$ to be $\mathcal{O}(10^{-2})$, which is the minimal requirement from the CMB observation, the probability to obtain a classical expanding universe is extremely small and negligible for sub-Planckian values of $V_0$, which is a fairly mild assumption.

\section{Conclusion}
\label{sec:Conclusion}

In this paper, we have revisited the Hartle-Hawking no-boundary proposal in the light of the Swampland conjectures such as the refined dS conjecture and the distance conjecture.  
It has turned out that the refined dS conjecture is more constraining than the distance conjecture. 
As is well known, these Swampland conjectures are in tension with inflation.  
When the parameters $c$ and $c'$ of the dS conjecture are $\mathcal{O}(1)$, even the presence of the classical solution (more precisely, the saddle-point solution) is not guaranteed.  
Motivated by the observational success of the inflationary paradigm, one may require that the dimensionless parameters in the conjectures  be somewhat smaller than $\mathcal{O}(1)$.  However, we have pointed out and confirmed that significantly more severe bounds like $c, c' \lesssim \mathcal{O}(10^{-11})$ (inequality~\eqref{constraints_c}), or more conservatively, $c \lesssim \mathcal{O}(10^{-5})$ (inequality~\eqref{conservative_constraints_c}),  must be imposed to solve the issue of the small probability for a classical expanding universe in the no-boundary proposal.  Unless such tiny numbers are reasonably explained, we conclude that the no-boundary proposal and the refined dS conjecture are not compatible with each other.

It should be mentioned that this incompatibility does not mean the inconsistency of the no-boundary proposal itself under the Swampland conjectures.  It simply means that it is extremely unlikely to produce universes like ours when we adopt both the no-boundary proposal and the Swampland conjectures. It is still possible to produce quantum universes in the no-boundary proposal without contradicting the Swampland conjectures. 

If we omit the refined dS conjecture, the no-boundary proposal and the distance conjecture can be compatible, but the dimensionless parameter $d$ in the inequality is constrained to be smaller than unity in the large-field models (see inequality~\eqref{constraint_d}).  This bound could be strengthened since we only require that the cutoff scale is higher than the Hubble scale during inflation.  
It is also worth noting that the obtained upper bound on $d$ is larger than the lower bound on $d$ suggested in Ref.~\cite{Andriot:2020lea} for $\alpha \lesssim 36$.

Our analyses are based on the specific potential~\eqref{flat-potential}, but we expect the qualitative features are similar in generic inflation models (see the discussion below inequality~\eqref{constraints_c}).  Potential exceptions that may realize the eternal inflation (to gain the volume factor) and allowed by the Swampland conjectures are the small-field hilltop and inflection-point inflation models.  Most of the evidence for the (refined) dS conjecture is based on the asymptotic behavior of the scalar.  The local feature of the potential such as a hilltop may not be severely constrained by any principles of quantum gravity.   However, an analysis in Ref.~\cite{Brahma:2019iyy} has shown that eternal inflation does not take place in the hilltop model in the perturbative regime when one imposes the refined dS conjecture. 

We have demonstrated the incompatibility of the volume-weight solution to the no-boundary proposal and the Swampland conjectures. If we accept the Swampland conjectures, the original no-boundary proposal has to be modified significantly.  On the other hand, if we stick to the no-boundary proposal, the refined dS conjecture  has to be weakened.

\section*{Acknowledgment}
We thank Oliver Janssen and Andrei Linde for useful discussions and comments. H.M.~would like to thank Kazunori Kohri and Fuminobu Takahashi for useful discussions. This work is supported by IBS under the project code, IBS-R018-D1.

\appendix

\section{Saddle-point solution for generic slow-roll potentials \label{sec:generic_slow-roll}}

In this appendix, we obtain the saddle-point solution for the no-boundary cosmology in the case of generic slow-roll potentials. 
 The analysis in this appendix is a generalization of that in Ref.~\cite{Lyons:1992ua}.

Let us consider an expansion of a generic slow-roll potential,  
\begin{align}
V(\phi) = V_* \left( 1 + \sqrt{2 \epsilon} \phi + \frac{1}{2} \eta \phi^2 + \cdots \right) \; ,
\end{align}
where $V_*$, $\sqrt{\epsilon}$ and $\eta$ are real constants, and dots represent higher order terms.  The $\epsilon = (V' / V)^2 /2 |_{\phi = 0}$ and $\eta = V'' /V |_{\phi = 0}$ are slow-roll parameters evaluated at the origin, which are different from those evaluated at the CMB scale.
It is always possible to redefine the origin of the field so that the real part of the scalar field vanishes at the South Pole ($\phi_\text{SP}^\text{R} = 0$), and this is the most convenient choice for numerical simulation at each point in the parameter space. 
However, it is convenient to keep $\phi_\text{SP}^\text{R}$ dependence in the following analysis.  For simplicity, we assume $V_0 \equiv V(\phi_\text{SP}) \simeq V(\phi_\text{SP}^\text{R}) \simeq V_*$.

Remember that the scalar field cannot have a large kinetic energy around the South Pole because of the regularity condition.  The slow-roll Euclidean equations of motion are
\begin{align}
&\left( \frac{\text{d} a } {\text{d} \tau } \right)^2 =  - \frac{a^2 V}{3} + \frac{1}{\ell^2} \;  , \\
& \frac{\text{d}^2 \phi}{\text{d} \tau^2} + 3 \frac{\text{d} a } {\text{d} \tau } \frac{\text{d} \phi } {\text{d} \tau }=  a V' \; .
\end{align}
Note that we cannot use $H \simeq \text{const.}$ around the South Pole in contrast to the standard inflationary studies because of the presence of the spatial curvature term. 

Around the origin, the solution is obtained as
\begin{align}
a \simeq  & \sqrt{\frac{3  }{ \ell ^2 V_0}} \sin \left( \sqrt{\frac{V_0}{3}} \tau  \right) \; , \\
\phi \simeq & \phi_\text{SP} + \frac{V_*}{8} \left( \sqrt{2 \epsilon} + \eta \phi_\text{SP} \right) \tau^2 \; .\end{align}
For  $|\tau| \ll \sqrt{3/V_0}$, the potential can be approximated by its South Pole value, $V\simeq V_0$, and the expression of $a$ is consequently simple as if the potential is a cosmological constant.

On the other hand, in the asymptotic Lorentzian region where the curvature term is negligible, the equations of motion in terms of $\mathcal{N} \equiv \ln a$ 
\begin{align}
\left(  \frac{\text{d} \mathcal{N} } {\text{d} \tau }  \right)^2 =&  - \frac{V_*}{3} \left( 1 + \sqrt{2\epsilon} \phi + \frac{1}{2} \eta \phi^2 \right) \; , \\
\frac{\text{d} \mathcal{N} } {\text{d} \tau } \frac{\text{d} \phi} {\text{d} \tau } =& \frac{V_*}{3} \left( \sqrt{2\epsilon} + \eta \phi \right) \; , 
\end{align}
are solved as follows.
In the following, we only keep terms up to the first order of the slow-roll parameters, $\mathcal{O}(\sqrt{\epsilon}, \, \eta)$.
$\frac{\text{d} \mathcal{N} } {\text{d} \tau }$ is obtained as $\frac{\text{d} \mathcal{N} } {\text{d} \tau } = - i \sqrt{\frac{V_*}{3}} \left( 1+ \sqrt{\epsilon/2} \phi + \frac{\eta}{4} \phi^2 \right)$, so 
\begin{align}
\frac{\text{d} \phi} {\text{d} \tau } = i  \sqrt{\frac{V_*}{3}} \left( \sqrt{2\epsilon}+ \eta \phi \right) \; .
\end{align}
This is integrated to give
\begin{align}
\phi \simeq & \left( \phi_\text{SP} + \frac{\sqrt{2\epsilon}}{\eta}   \right) e^{i \sqrt{\frac{V_*}{3}} \eta \tau}  - \frac{\sqrt{2\epsilon}}{\eta} \nonumber \\
\simeq & \phi_\text{SP} + i \sqrt{\frac{V_*}{3}} \left( \sqrt{2\epsilon}+ \eta \phi_\text{SP} \right) \tau \; .
\end{align}
Up to the first order, $\frac{\text{d} \mathcal{N} } {\text{d} \tau } = - i \sqrt{\frac{V_0}{3}}$, so 
\begin{align}
a \simeq   i  \sqrt{\frac{3  }{ \ell ^2 V_0}}   \exp \left( - i \sqrt{\frac{V_0}{3}} \tau \right) \; .
\end{align}
Note that we cannot consider too long a period for $\Delta \tau$ since the slow-roll suppressed terms become non-negligible at some point. 

$\phi$ is real on the line $\tau^\text{R} =  - \sqrt{3/V_0} \phi_\text{SP}^\text{I} /(\sqrt{2\epsilon} + \eta \phi_\text{SP}^\text{R})$. Here and hereafter, we regard $\phi_\text{SP}^\text{I}$ as a first-order quantity and neglect terms such as $\mathcal{O}(\eta \phi_\text{SP}^\text{I})$.  $a$ is real on the same line if and only if 
\begin{align}
\phi_\text{SP}^\text{I} \simeq - \frac{\pi}{2} (\sqrt{2\epsilon} + \eta \phi_\text{SP}^\text{R}) \; .
\end{align}
This shows consistency of the assumption $\phi_\text{SP}^\text{I} = \mathcal{O}(\sqrt{\epsilon}, \eta )$.
Alternatively, we may look at the reality condition for $a$ first.  Then, we can see that it is always given by $\tau^\text{R} = \sqrt{3/V_0} \pi/2$.  Then, the condition for the reality of $\phi$ is
\begin{align}
\phi_\text{SP}^\text{I} \simeq \left( \frac{\sqrt{2\epsilon}}{\eta} + \phi_\text{SP}^\text{R}\right) \tan \left( - \frac{\pi}{2} \eta \right).
\end{align}
This is consistent with the above expression of $\phi_\text{SP}^\text{I}$ when the slow-roll parameters are small.

Using these formulae, the asymptotic solutions are
\begin{align}
\phi \simeq &  \phi_\text{SP}^\text{R} - \sqrt{\frac{V_*}{3}} (\sqrt{2\epsilon} + \eta \phi_\text{SP}^\text{R}) t \; , \\
a \simeq & \sqrt{\frac{3}{\ell^2 V_0^\text{R}}} \exp \left(\sqrt{\frac{V_0^\text{R}}{3}} t \right) \; ,
\end{align}
where $\tau = - \sqrt{3/V_*} \phi_\text{SP}^\text{I} / (\sqrt{2\epsilon} + \eta \phi_\text{SP}^\text{R})+ i t $.

The approximate inverse functions of these are 
\begin{align}
t \simeq & \sqrt{\frac{3}{V(\chi)}} \ln \left(  \sqrt{\frac{\ell^2 V (\chi) }{3} }b \right) \; , \\
\phi_\text{SP}^\text{R} \simeq & \chi \left( 1 + \eta  \ln \left(  \sqrt{\frac{\ell^2 V (\chi)}{3} }b \right) \right) + \sqrt{2 \epsilon} \ln \left(  \sqrt{\frac{\ell^2 V(\chi)}{3} }b \right) \; .
\end{align}

The Euclidean action is evaluated as follows
\begin{align}
S_\text{E} \simeq &\,  4\pi^2 \int_0^{\frac{\pi}{2}\sqrt{\frac{3}{V_*}}} \text{d} \tau \left( (\ell a)^3 V_0 -  3 \ell a \right) + 
    4 \pi^2 i \int_0^{t} \text{d}\tau^\text{I}  (\ell a)^3 V(\phi)   \nonumber \\
    \simeq &  - \frac{12 \pi^2}{V_0} + \frac{12 i \pi^2}{V_*} e^{3 \sqrt{\frac{V_*}{3}}t} \left( 1- \phi_\text{SP}^\text{R} \left( \frac{3}{2} \sqrt{\frac{V_*}{3}} t - 1\right) \left( \sqrt{2 \epsilon} + \frac{\eta}{2} \phi_\text{SP}^\text{R}  \right)  \right) \nonumber \\
\simeq & - \frac{12 \pi^2}{V(\phi_\text{SP}(\chi, b))} + \frac{4 \sqrt{3}\pi^2 i }{3} V(\chi)^{1/2} (\ell b)^3 \; .
\end{align}
In this way, the real part of the Euclidean action is determined essentially by the South Pole value of the potential.  The probability~\eqref{probability} is evaluated as
\begin{align}
P(\chi, b ) \approx \exp \left( \text{Re} \, \frac{24 \pi^2}{V(\phi_\text{SP} (\chi, b) )} \right) \; ,
\end{align}
and this is larger for a smaller value of the potential $V$.

The derivatives of the real and imaginary parts of the Euclidean action are
\begin{align}
(\nabla S_\text{E}^\text{R})^2 =&  - \frac{1}{12 \pi^2 \ell^3 b} (\partial_b S_\text{E}^\text{R})^2 + \frac{1}{2 \pi^2 \ell^3 b^3} (\partial_\chi S_\text{E}^\text{R})^2   \nonumber \\
   =& \mathcal{O}(\epsilon, \sqrt{\epsilon}\eta, \eta^2)  +  \mathcal{O}(\epsilon, \sqrt{\epsilon}\eta, \eta^2)  \; , \\
   (\nabla S_\text{E}^\text{I})^2 =&  - \frac{1}{12 \pi^2 \ell^3 b} (\partial_b S_\text{E}^\text{I})^2 + \frac{1}{2 \pi^2 \ell^3 b^3} (\partial_\chi S_\text{E}^\text{I})^2   \nonumber \\
   =& 4 \pi^2 (\ell b)^3 V(\chi) + 
   \mathcal{O}(\epsilon, \sqrt{\epsilon}\eta, \eta^2) \; .
\end{align}
From these expressions, $|\partial_b S_\text{E}^\text{R} | \ll |\partial_b S_\text{E}^\text{I} | $ clearly follows. Also, it is easy to see $|\partial_\chi S_\text{E}^\text{R} | / |\partial_\chi S_\text{E}^\text{I} |  \sim | S_\text{E}^\text{R} / S_\text{E}^\text{I}|\sim 1/(V^{3/2} (\ell b)^3)$. 
Thus, the classicality condition~\eqref{WKB} 
 is satisfied better for larger $V(\chi)$ and larger $b$.   As expected, the results in this subsection are qualitatively similar to the cosmological constant case reviewed in sec.~\ref{sec:example}.

\section*{Note added in proof}
Some minor errors in Appendix~\ref{sec:generic_slow-roll} were pointed out by Oliver Janssen, which have been corrected in this version. For a detailed discussion of the slow-roll approximation in quantum cosmology and its difference from the standard cosmological case, see Ref.~\cite{Janssen:2020pii}.

\nocite{}
\bibliographystyle{JHEP}
\bibliography{ref}

\end{document}